\documentclass[twocolumn]{aastex61}
\usepackage{xspace} 
\usepackage{graphicx} 
\usepackage{natbib}
\usepackage{amsmath}
\newcommand{\gsim}{\ensuremath{\,\gtrsim\,}\xspace}
\newcommand{\lsim}{\ensuremath{\,\lesssim\,}\xspace}
\newcommand{\vlsr}{\ensuremath{V_{\rm LSR}}\xspace}
\newcommand{\kms}{\ensuremath{\,{\rm km\,s^{-1}}}\xspace}
\newcommand{\hi}{H\,{\sc i}}
\newcommand{\hii}{H\,{\sc ii}}
\newcommand{\khz}{\ensuremath{\,{\rm kHz}}\xspace}
\newcommand{\mhz}{\ensuremath{\,{\rm MHz}}\xspace}
\newcommand{\ghz}{\ensuremath{\,{\rm GHz}}\xspace}
\revised{\today}
\submitjournal{\apjs\ (accepted for publication)}

\shorttitle{SHRDS I: Bright Catalog}
\shortauthors{Wenger et al.}

\begin{document}

\title{The Southern \hii\ Region Discovery Survey I: The Bright Catalog}

\author{Trey V. Wenger} 
\affiliation{Astronomy Department, University of Virginia, P.O. Box
  400325, Charlottesville, VA 22904-4325, USA.}
\affiliation{National Radio Astronomy Observatory, 520 Edgemont Road,
  Charlottesville, VA 22903, USA.}
\email{tvw2pu@virginia.edu}

\author{John. M. Dickey}
\affiliation{School of Natural Sciences, University of Tasmania,
  Hobart, TAS 7001, Australia.}

\author{C. H. Jordan}
\affiliation{International Centre for Radio Astronomy Research,
  Curtin University, Bentley, WA 6102, Australia.}
\affiliation{ARC Centre of Excellence for All Sky Astrophysics in 3
  Dimensions (ASTRO 3D), Curtin University, Bentley 6845, Australia}

\author{Dana S. Balser}
\affiliation{National Radio Astronomy Observatory, 520 Edgemont Road,
  Charlottesville, VA 22903, USA.}

\author{W. P. Armentrout}
\affiliation{Department of Physics and Astronomy, West Virginia
  University, Morgantown, WV 26505, USA.}
\affiliation{Center for Gravitational Waves and Cosmology, West
Virginia University, Morgantown, Chestnut Ridge Research Building, 
Morgantown, WV 26505, USA.}
\affiliation{Green Bank Observatory, P.O. Box 2, Green Bank, WV 24944, USA.}

\author{L. D. Anderson}
\affiliation{Department of Physics and Astronomy, West Virginia
  University, Morgantown, WV 26505, USA.}
\affiliation{Center for Gravitational Waves and Cosmology, West
Virginia University, Morgantown, Chestnut Ridge Research Building, 
Morgantown, WV 26505, USA.}
\affiliation{Adjunct Astronomer at the Green Bank Observatory, 
P.O. Box 2, Green Bank, WV 24944, USA.}

\author{T. M. Bania}
\affiliation{Institute for Astrophysical Research, Astronomy 
  Department, Boston University, 725 Commonwealth Ave., Boston, MA 
  02215, USA.}

\author{J. R. Dawson}
\affiliation{Department of Physics and Astronomy and MQ Research
  Centre in Astronomy, Astrophysics, and Astrophotonics,
  Macquarie University, NSW 2109, Australia.}
\affiliation{Australia Telescope National Facility, CSIRO Astronomy
  and Space Science, P.O. Box 76, Epping, NSW 1710, Australia.}

\author{N. M. McClure--Griffiths}
\affiliation{Research School of Astronomy and Astrophysics,
  Australian National University, Canberra, ACT 2611, Australia.}

\author{Jeanine Shea}
\affiliation{National Radio Astronomy Observatory, 520 Edgemont Road,
  Charlottesville, VA 22903, USA.}
\affiliation{Department of Physics and Astronomy, Bucknell University,
  153 Olin Science Building, Lewisburg, PA 17837, USA.}

\begin{abstract}
  The census of Galactic \hii\ regions is vastly incomplete in the
  Southern sky. We use the Australia Telescope Compact Array (ATCA) to
  observe 4--10\ghz\ radio continuum and hydrogen radio recombination
  line (RRL) emission from candidate \hii\ regions in the Galactic
  zone \(259^\circ < \ell < 344^\circ, |b| < 4^\circ\). In this first
  data release, we target the brightest \hii\ region candidates and
  observe 282 fields in the direction of at least one previously-known
  or candidate \hii\ region. We detect radio continuum emission and
  RRL emission in 275 (97.5\%) and 258 (91.5\%) of these fields,
  respectively. We catalog the \({\sim}7\ghz\) radio continuum peak
  flux densities and positions of 80 previously-known and 298
  candidate \hii\ regions. After averaging \({\sim}18\) RRL
  transitions, we detect 77 RRL velocity components towards 76
  previously-known \hii\ regions and 267 RRL velocity components
  towards 256 \hii\ region candidates. The discovery of RRL emission
  from these nebulae increases the number of known Galactic
  \hii\ regions in the surveyed zone by 82\%, to 568 nebulae. In the
  fourth quadrant we discover 50 RRLs with positive velocities,
  placing those sources outside the Solar circle. Including the pilot
  survey, the SHRDS has now discovered 295 Galactic \hii\ regions. In
  the next data release we expect to add \({\sim}\)200 fainter and
  more distant nebulae.
\end{abstract}

\keywords{Galaxy: kinematics and dynamics -- Galaxy: structure --
  (ISM:) \hii\ regions -- ISM: kinematics and dynamics -- radio lines:
  ISM -- surveys}

\section{Introduction}

Massive OB-type stars ionize the natal gas in their surroundings,
creating \hii\ regions. Since these nebulae have short lifetimes
(\(\lsim\) 10 Myr) they are the locations of \textit{current}
high-mass star formation in the Galaxy. \hii\ regions are the classic
tracer of Galactic spiral structure, and their chemical abundances
reveal the metallicity of the interstellar medium (ISM) in which they
formed. A complete census of Galactic \hii\ regions would inform
models of both Galactic kinematics as well as the formation and
chemo-dynamical evolution of the Galaxy.

More than 60 years ago, \citet{sharpless1953} and
\citet{sharpless1959} began surveys of Galactic \hii\ regions.
Starting with photographic plates from the 48-inch Schmidt telescope
(now known as the Samuel Oschin Telescope) at the Palomar Observatory,
\citet{sharpless1953} compiled a catalog of 142 ``emission nebulae''
and stars associated with those nebulae. \citet{sharpless1959}
expanded upon his previous work using the newly-completed National
Geographic-Palomar Sky Atlas.  Adopting the term ``\hii\ region'' from
\citet{stromgren1948}, this second catalog contains 313 optical
\hii\ regions covering the entire sky north of declination
\(-27^\circ\). \citet{gum1955}, and later \citet{rodgers1960},
expanded the \hii\ region survey to the Southern hemisphere using
H\(\alpha\) photographic plates from the Mount Stromolo Observatory.

\begin{figure}
  \centering
  \includegraphics[width=\linewidth]{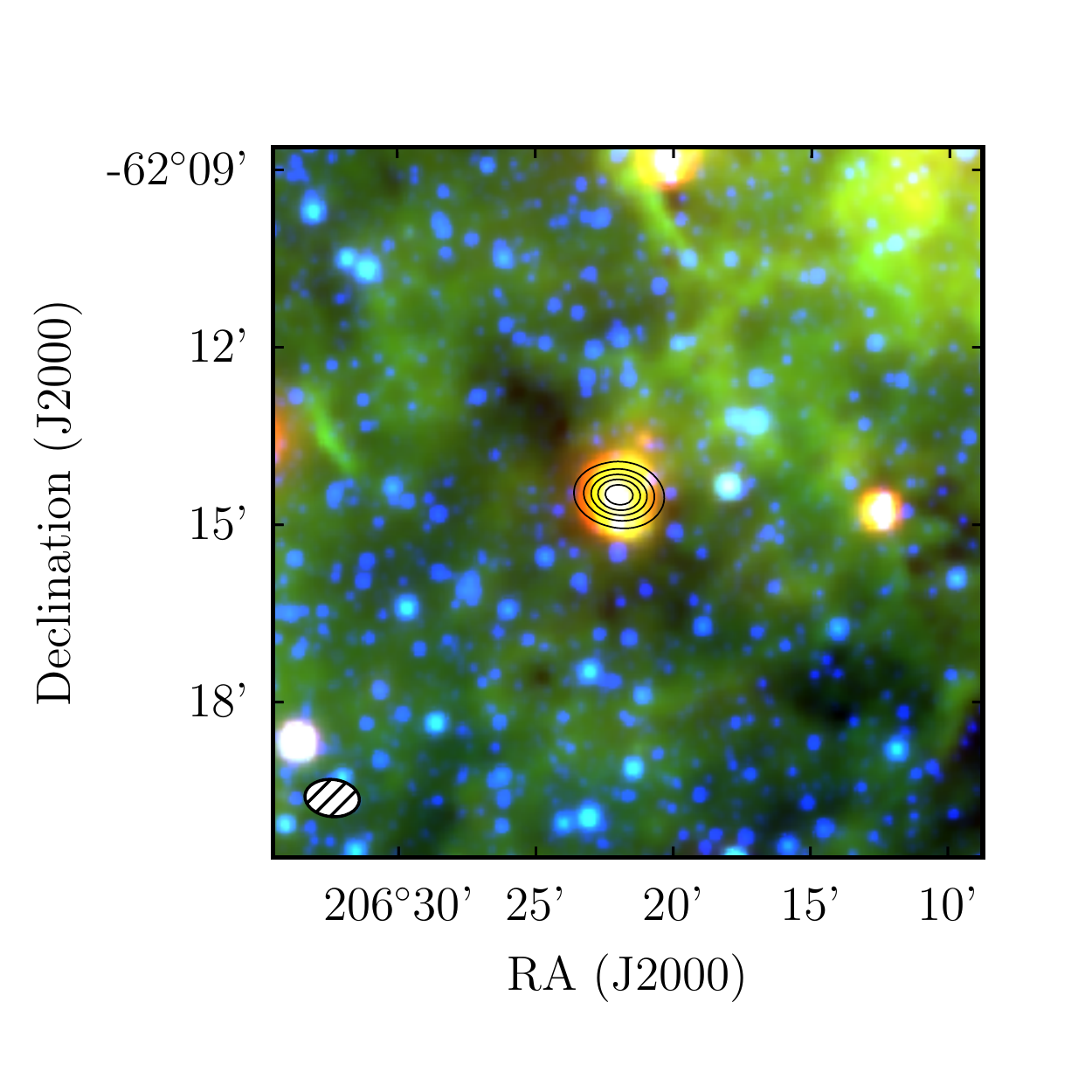}
  \caption{Infrared image of a typical \hii\ region candidate from the
    \textit{WISE} Catalog, G309.176$-$00.028. The image is a composite
    of \textit{WISE} 22\(\mu\)m (red), 12\(\mu\)m (green), and
    3.4\(\mu\)m (blue) data. The black contours are the SHRDS
    7\ghz\ continuum emission (50 mJy beam\(^{-1}\) to 250 mJy
    beam\(^{-1}\) in 50 mJy beam\(^{-1}\) intervals). The hatched
    ellipse represents the ATCA half-power synthesized beam.}
  \label{fig:shrds386}
\end{figure}

With the prediction of radio recombination lines (RRLs) by
\citet{kardashev1959} and their subsequent discovery by
\citet{hoglund1965a,hoglund1965b}, there was now an extinction-free
spectroscopic tracer of optically obscured \hii\ regions.  The first
generation of RRL \hii\ region surveys was carried out by
\citet{reifenstein1970} and \citet{wilson1970}. Using the National
Radio Astronomy Observatory (NRAO; now Green Bank Observatory) 140
Foot telescope, \citet{reifenstein1970} detected the H109\(\alpha\)
RRL toward 82 Galactic \hii\ regions. \citet{wilson1970} extended the
survey to the Southern sky using the NRAO 6 cm receiver on the
210-foot Parkes Telescope. They detected H109\(\alpha\) RRL emission
toward 130 Galactic \hii\ regions, bringing the total census to 212
nebulae with RRL detections. These projects were successful despite
the low spectral resolution (\({\sim}6\,\text{km s\(^{-1}\)}\)) and
high system temperatures (\({\sim}100\) K) of their instruments.

\begin{figure*}
  \centering
  \includegraphics[width=\linewidth]{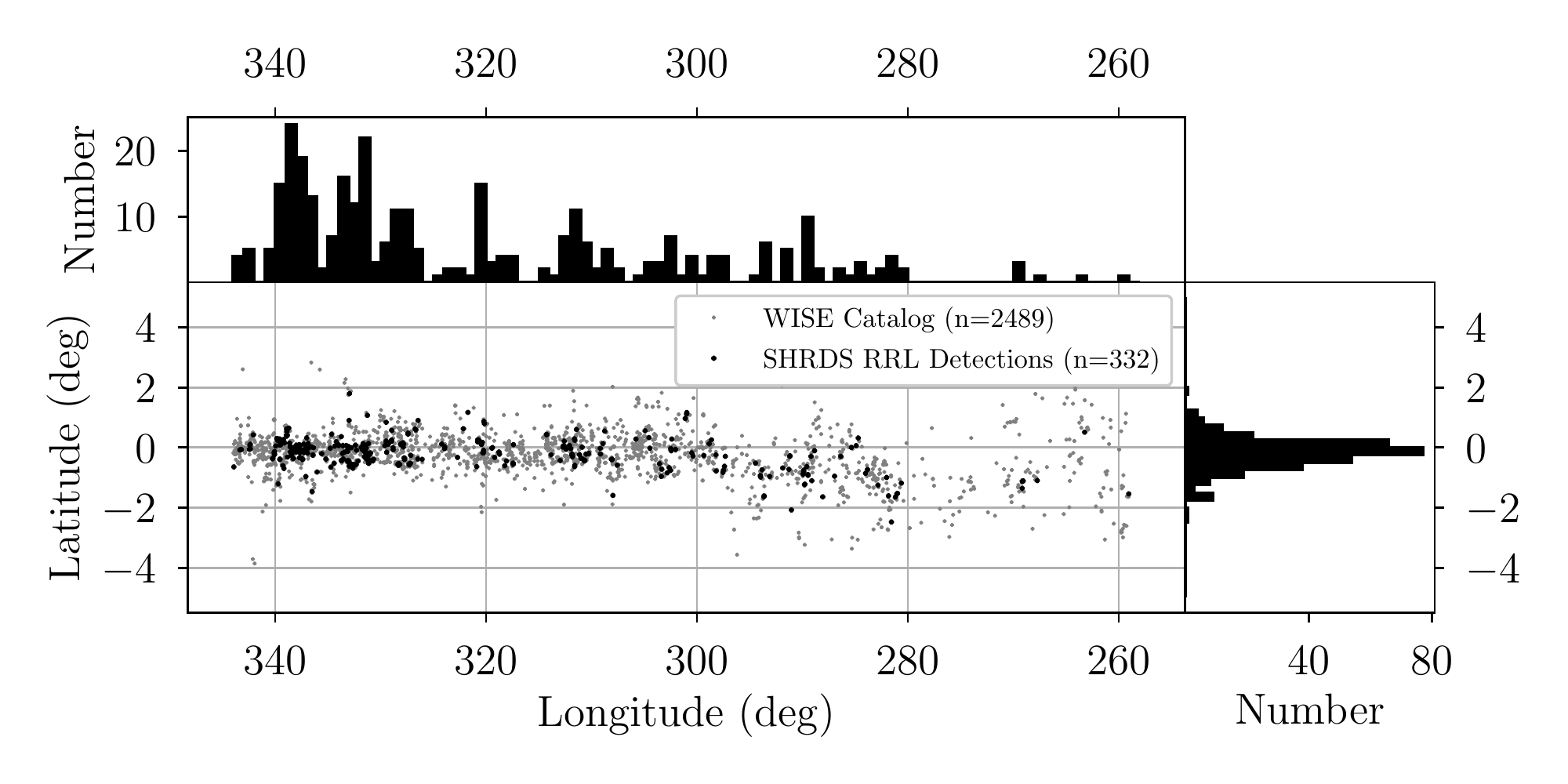}
  \caption{The Galactic positions of all \textit{WISE} Catalog
    \hii\ regions and \hii\ region candidates with \(259^\circ < \ell
    < 344^\circ\), \(|b| < 4^\circ\) (gray points) as well as the
    subset of those observed in SHRDS fields (black points). The
    histograms show the Galactic distribution of the observed SHRDS
    targets.}
  \label{fig:positions}
\end{figure*}

Equipped with better telescopes, more sensitive receivers, and more
advanced correlators, the second generation of surveys began with
\citet{downes1980} in the Northern sky and \citet{caswell1987} in the
Southern sky. Using the Effelsburg 100-m telescope, \citet{downes1980}
targeted 262 bright sources from the recently-completed
5\ghz\ continuum survey by \citet{altenhoff1979} in search of
H110\(\alpha\) RRL emission and H\(_2\)CO absorption. They detected
RRL emission toward 171 nebulae. \citet{caswell1987} used the updated
1024-channel digital correlator on the Parkes telescope to
simultaneously observe two RRLs (H109\(\alpha\) and H110\(\alpha\)) as
well as H\(_2\)CO. By averaging the two RRL transitions, they were
able to detect RRL emission from 316 Galactic \hii\ regions. Using the
NRAO 140 Foot telescope, \citet{lockman1989} observed all remaining
reasonably bright compact radio continuum sources (\(\gsim\) 1 Jy
beam\(^{-1}\)) in the \citet{altenhoff1979} survey. This generation of
RRL surveys was completed by \citet{lockman1996} who observed faint
and diffuse radio sources in search of angularly large
\hii\ regions. These 140 Foot telescope surveys discovered
approximately 350 new Galactic \hii\ regions, bringing the total
census of known Galactic \hii\ regions to about 1000 nebulae.

With the completion of the \citet{lockman1989} and \citet{lockman1996}
surveys, all of the bright radio continuum sources in the
\citet{altenhoff1979} catalog were observed, and systematic searches
for Galactic \hii\ regions ceased. It was apparent, however, that the
census of \hii\ regions was vastly incomplete; only a handful of
\hii\ regions were known in the outer Galaxy in the first and fourth
Galactic quadrants, for example. A deeper RRL survey would discover
fainter and more distant \hii\ regions, allowing us to explore both
Galactic structure and the properties of high-mass star formation
beyond the Galactic Center.

\textit{We are now completing the third generation of RRL surveys of
  \hii\ regions.} This generation is motivated by two great
advancements in the field: 1) deep all-sky infrared surveys, which are
sensitive to the thermal dust emission associated with \hii\ regions
across the Galactic disk, and 2) ultra-sensitive radio telescopes with
wide-bandpass receivers and correlators, which can simultaneously
observe many RRL transitions. The Green Bank Telescope \hii\ Region
Discovery survey \citep[GBT HRDS;][]{bania2010,anderson2011} used the
largest fully-steerable telescope in the world to discover 448 new
Galactic \hii\ regions in the first and second Galactic
quadrants. Follow-up surveys with the GBT and the Arecibo Telescope
\citep{bania2012,anderson2015b,anderson2018} added another 439
\hii\ regions, bringing the total number of HRDS discoveries to
887. These surveys more than doubled the number of known \hii\ regions
in the surveyed zone and completed the census of Northern sky
\hii\ regions brighter than \({\sim}100\,\text{mJy beam\(^{-1}\)}\) at
\({\sim}9\ghz\).

The Southern \hii\ Region Discovery Survey (SHRDS) is an extension of
the HRDS into the Southern sky. Using the sensitive and wide-bandpass
receivers on the Australia Telescope Compact Array (ATCA), we aim to
complete the census of Southern sky \hii\ regions to nearly the same
sensitivity limit as the HRDS (\({\sim}100\,\text{mJy
  beam\(^{-1}\)}\)). At the conclusion of the SHRDS, we will have a
catalog of all Galactic \hii\ regions ionized by at least a single
O-star. This catalog will reveal new insights into the current
structure, formation, and evolutionary history of the Milky Way.

\section{Target Sample}

Our targets are selected from the \textit{Wide-field Infrared Survey
  Explorer (WISE)} Catalog of Galactic \hii\ Regions
\citep{anderson2014}. The \textit{WISE} Catalog is the most complete
census of known and candidate \hii\ regions extant.  Candidate
\hii\ regions are identified based on their spatially coincident 12
\(\mu\)m, 22 \(\mu\)m, and, if available, radio emission. The infrared
data are taken from the \textit{WISE} All-Sky data \citep{wright2010}
and the radio data from MAGPIS \citep{becker1994,helfand2006}, VGPS
\citep{stil2006}, CGPS \citep{taylor2003}, NVSS \citep{condon1998},
SGPS \citep{mccluregriffiths2005}, and SUMSS
\citep{bock1999,mauch2003}. The 12 \(\mu\)m and 22 \(\mu\)m emission
stems from the polycyclic aromatic hydrocarbons (PAHs) in the
photodissociation region (PDR) surrounding the \hii\ region and the
warm dust associated with the \hii\ region, respectively. The radio
emission is caused by thermal (free-free) emission from the ionized
gas. The \textit{WISE} Catalog contains about 8000 objects:
\({\sim}2000\) known \hii\ regions, \({\sim}2000\) radio-loud
\hii\ region candidates, and \({\sim}4000\) radio-quiet \hii\ region
candidates. Radio-quiet \hii\ region candidates are sources that have
not been detected in existing radio continuum surveys.
Figure~\ref{fig:shrds386} shows the \textit{WISE} infrared image and
SHRDS observed 7\ghz\ radio continuum contours for the \hii\ region
candidate, G309.176$-$00.028.

\begin{figure}
  \centering
  \includegraphics[width=\linewidth]{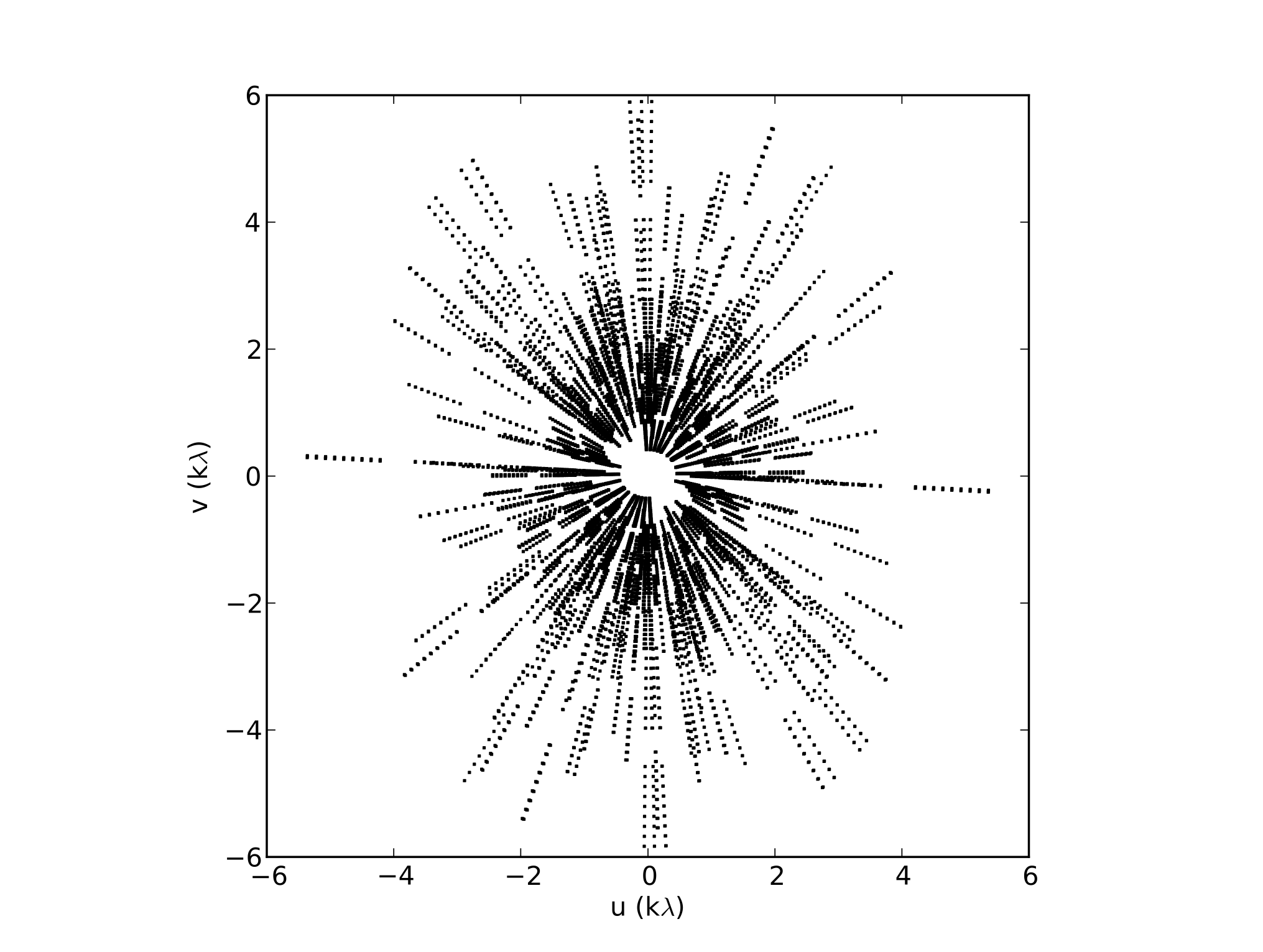}
  \caption{A representative example of the \textit{uv}-coverage
    obtained for each SHRDS field. For clarity, only 9 of the 66
    observed continuum window channels are shown. These data are the
    combination of 30 \({\sim}2\) minute snapshots split equally
    between the H75 and H168 antenna configurations.}
  \label{fig:uv_coverage}
\end{figure}

In this first data release, we target \hii\ region candidates in the
Galactic longitude range \(259^\circ < \ell < 344^\circ\) with a
predicted 6\ghz\ peak flux density greater than 60 mJy beam\(^{-1}\).
This longitude range contains the portion of the sky visible by the
ATCA that could not be observed by the HRDS telescopes.  We estimate
the 6\ghz\ peak flux density by extrapolating from the measured SUMSS
843\mhz\ flux density assuming \(S_{\rm 6\ghz}/S_{\rm 843\mhz} = ({\rm
  6\ghz}/{\rm 843\mhz})^\alpha\) with an optically thin spectral index
of \(\alpha = -0.1\). We also observe several previously known
\hii\ regions from the GBT HRDS, \citet{caswell1987}, and
\citet{wilson1970} catalogs. These observations allow us to test our
data reduction and analysis procedure and to compare single-dish and
interferometric results.

Our target list for the SHRDS Bright Catalog contains 257 \hii\ region
candidates and 25 previously-known \hii\ regions. In some cases,
multiple targets are observable within one ATCA primary beam, so we
group them into a single pointing, a ``field,'' centered between the
targets. We observe 282 individual fields that contain many more
\textit{WISE} Catalog sources than the 282 bright targets. In total,
there are 632 \hii\ region candidates and 149 previously-known
\hii\ regions within our fields, but most of these will be too distant
and faint or too large and diffuse to be detected in our survey.
Table~\ref{tab:target_sample} lists information about the
\hii\ regions and \hii\ region candidates in each field. Listed for
each field is the field name, center position, and observing
epoch. Each field contains multiple sources, and for each source we
list the \textit{WISE} Catalog source name, \textit{WISE} Catalog
designation (K for previously-known \hii\ region, C for \hii\ region
candidate, Q for radio-quiet \hii\ region candidate, and G for
\hii\ region candidate associated with a group of \hii\ regions),
\textit{WISE} infrared position, \textit{WISE} infrared radius,
\(R_{\rm IR}\), predicted 6\ghz\ peak continuum flux density, \(S_{\rm
  6 GHz}\), the separation between the position of the SUMSS continuum
peak emission and the infrared position, \(\Delta \theta\), and the
reference to the previously-known RRL detection, if any.  A
superscript ``T'' on the source name indicates that this object meets
our ``nominal'' target criteria: a predicted 6\ghz\ peak continuum
flux density brighter than 60 mJy beam\(^{-1}\) and a predicted radio
diameter smaller than the maximum recoverable scale of the ATCA at
6\ghz, \({\sim}265\) arcseconds. We estimate that the radio diameter
is half of the \textit{WISE} Catalog infrared diameter
\citep[e.g.][]{bihr2016}. There are 179 \hii\ region candidates and
100 previously-known \hii\ regions in our fields meeting our nominal
criteria. About 6\% of our fields do not contain a nominal target due
to a larger size criterion in our early observations.
Figure~\ref{fig:positions} shows the positions of all \textit{WISE}
Catalog \hii\ regions and \hii\ region candidates with \(259^\circ <
\ell < 344^\circ\), as well as the subset of those observed in SHRDS
fields.

\section{Observations}

We use the ATCA to observe radio continuum and hydrogen RRL emission
in each field. The observing procedure and correlator configuration
are similar to that used in the SHRDS pilot project
\citep{brown2017}. Data included here were observed June - October
2015 and July - September 2016. In total we observe 478 hours split
nearly equally between the most compact H75 antenna configuration and
the more extended H168 antenna configuration. A summary of the
observing dates, hours observed, and antenna configurations is given
in Table~\ref{tab:observations}.

The C/X-band receiver on the ATCA covers \(4-6\ghz\) with a
\({\sim}20\) K system temperature. The Compact Array Broadband Backend
(CABB) simultaneously measures both low spectral resolution, large
bandwidth radio continuum spectral windows (hereafter, continuum
windows) as well as many high spectral resolution, small bandwidth
spectral windows (hereafter, spectral line windows). We use CABB in
the 64\mhz\ mode, which allows us to simultaneously observe two
2\ghz\ bandwidth continuum windows (4.5--6.5\ghz\ and 7.5--9.5\ghz)
and thirty-two 64\mhz\ bandwidth spectral line windows. The continuum
windows have 33 channels (64\mhz/channel) and the spectral line
windows have 2048 channels (31.25\khz/channel). We tune the spectral
line windows to 20 different hydrogen RRLs, as summarized in
Table~\ref{tab:backend}. For each spectral window we list the center
frequency, \(\nu_{\rm center}\), bandwidth, number of channels, and channel
width, \(\Delta\nu\). For the spectral line windows we also identified
the targeted RRL and RRL rest frequency, \(\nu_{\rm RRL}\). The data
are observed in two circular polarizations, LL and RR, thus yielding
40 independent RRL spectra.

Our spectral window configuration is a compromise between the two
setups used in the pilot survey. At lower frequencies, the RRLs are
more tightly spaced in frequency and we can observe more of them
within a given bandwidth, but these RRLs are fainter and contaminated
with more radio frequency interference (RFI). At higher frequencies,
the RRLs are brighter, but they are more spaced out in frequency and
thus we observe fewer of them within a given bandwidth.

We use a two-phase observing strategy for the SHRDS: a first-pass
``snapshot'' survey in the H75 antenna configuration to measure the
continuum brightness of each target, and then a second-pass ``deep''
survey split between the H75 and H168 antenna configurations to
measure the RRLs. In the ``snapshot'' survey we observe each target
for a total of \({\sim}20\) minutes in \({\sim}2-3\) minute
integrations spread over \({\sim}9\) hours in hour angle. We reduce
these data and measure the continuum flux density of each target.
Assuming a RRL-to-continuum intensity ratio of \(0.10\), typical for
an optically thin \hii\ region in local thermodynamic equilibrium
(LTE) at 6\ghz \citep[e.g.][]{wilson1970,lockman1975}, we estimate the
integration time required to detect the RRL after averaging the 20
observed RRL transitions. We also take note of the level of confusion
in each field and we give confused fields priority in the H168 antenna
configuration. In the ``deep'' survey, we re-observe each target for
this estimated integration time split between the H75 and H168 antenna
configurations.

\begin{figure}
  \centering
  \includegraphics[width=\linewidth]{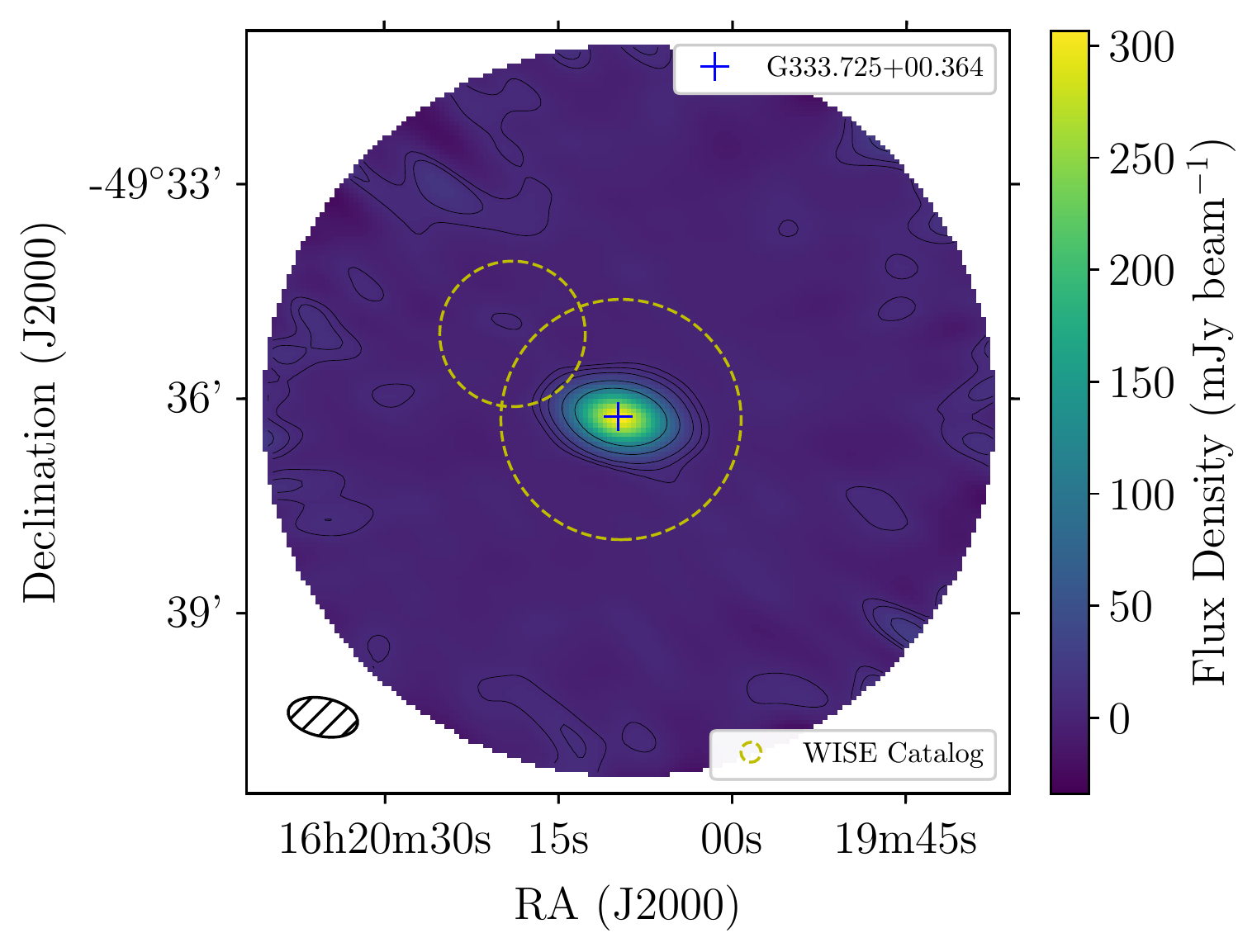} \\
  \includegraphics[width=\linewidth]{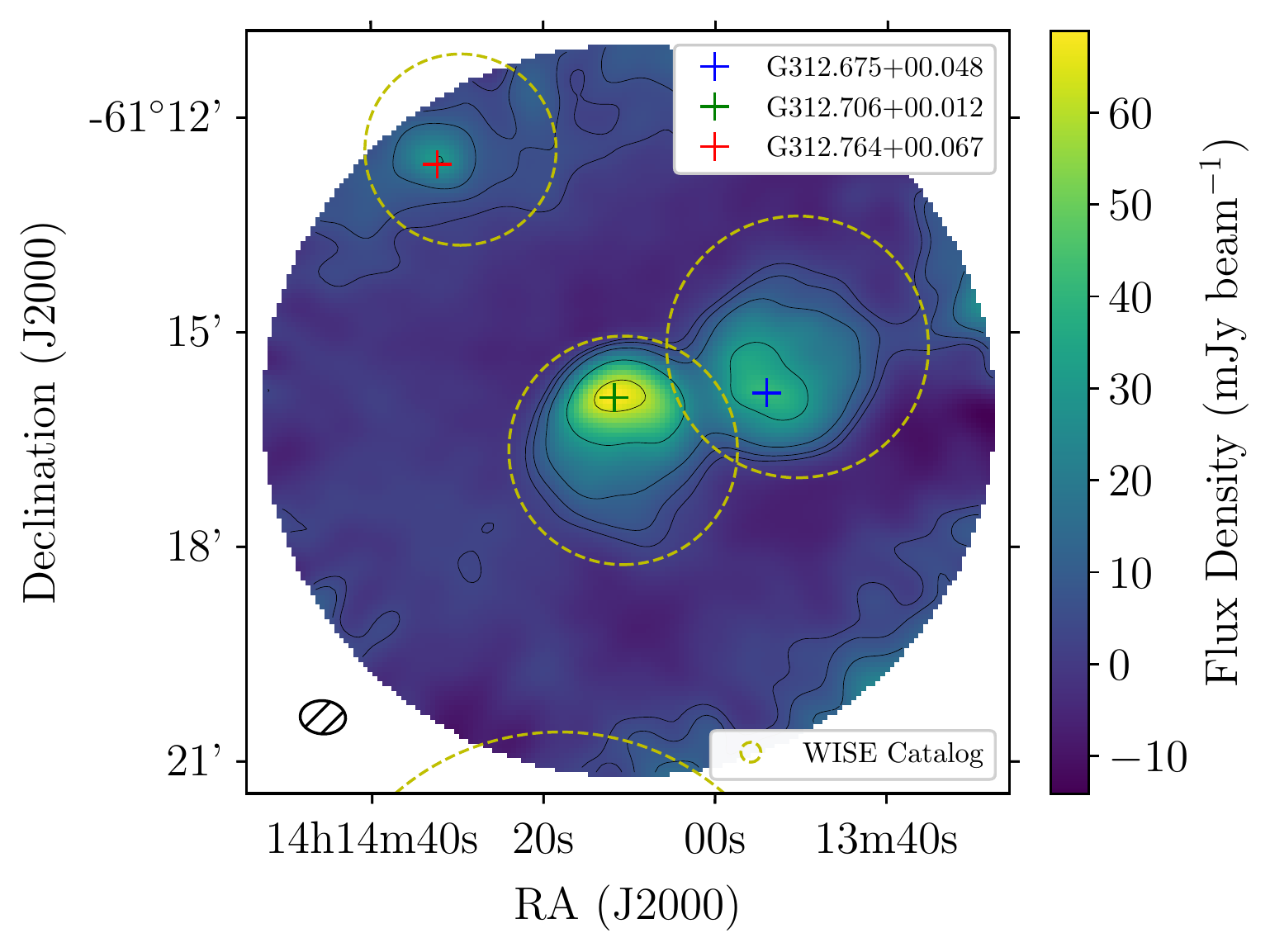}
  \caption{Representative 8--10\ghz\ continuum band images for each
    continuum quality factor (QF). The top panel is field ``shrds803''
    (non-tapered), which contains a continuum QF A target,
    G333.725+00.364 (un-resolved, un-confused, and centered). The
    bottom panel is field ``shrds462'' (non-tapered), which contains
    two continuum QF B tagets, G312.706+00.012 and G312.675+00.048
    (resolved and spatially confused), as well as a QF C target,
    G312.764+00.067 (resolved and far off-center). The black contours
    are at 5, 10, 20, 50, and 100 times the field center rms
    noise. The yellow dashed circles represent infrared positions and
    sizes of the objects in the \textit{WISE} Catalog, the crosses are
    the locations of the peak continuum emission, and the hatched
    ellipse represents the ATCA half-power synthesized beam at 7\ghz.}
  \label{fig:cont_qf}
\end{figure}

Unlike the pilot survey, we aim to create high dynamic range images of
our targets. Therefore, we require longer integrations and good
coverage in the \textit{uv-}plane. Total integration times on our
targets range from \({\sim}30\) minutes to \({\sim}60\) minutes spread
over \({\sim}9\) hours in hour angle depending on the predicted RRL
intensity. Figure~\ref{fig:uv_coverage} shows an example of the
typical \textit{uv-}coverage for our observations. Our observing
strategy nearly fills the \textit{uv-}plane and yields high-fidelity,
high-dynamic range images.  Due to the nature of an interferometer,
however, we are not sensitive to any emission on spatial scales larger
than the maximum recoverable scale of our most compact antenna
configuration, \({\sim}265\) arcseconds at 6\ghz.  Since over 95\% of
the sources in the \textit{WISE} Catalog have predicted radio sizes
smaller than 265 arcseconds, the ATCA is optimized for the SHRDS.

\vspace{0.4cm}

\section{Data Reduction and Analysis}

We develop a publicly-available, modular data calibration, reduction,
and analysis pipeline for the SHRDS, the Wenger Interferometry
Software Package
\citep[WISP;][]{wisp}\footnote{\url{https://doi.org/10.5281/zenodo.2225273}}. The
pipeline is written in \textit{Python} and is implemented through the
\textit{Common Astronomy Software Applications} package
\citep[CASA;][]{mcmullin2007}. These tools are generic and may be used
to reduce and analyze \textit{any} continuum or spectral line
interferometric data set. They benefit from a balance of automation
and user-input with tunable parameters to handle multiple
use-cases. For example, the calibration pipeline uses built-in CASA
automatic bad-data flagging algorithms, but also generates many data
quality diagnostics used for manual flagging. Here we briefly describe
the specific calibration, reduction, and analysis steps used for the
SHRDS. A more complete discussion of WISP and our data reduction
process is in Appendix~\ref{sec:app_data_reduction}.

\subsection{Calibration}

We observe at least one ``primary'' calibrator and several
``secondary'' calibrators each day. The primary calibrators, 0823--500
and 1934--638, are used to calibrate the absolute flux and delays and
to remove instrumental bandpass structure. The secondary calibrators
are point sources located close to our science targets on the sky and
are used to calibrate the complex gains (phases and amplitudes) of our
data. We typically observe a primary calibrator twice during a single
observing session and a secondary calibrator every 10--20
minutes. Table~\ref{tab:observations} lists the calibrators we used in
the Bright Catalog.

\begin{figure*}
  \centering
  \includegraphics[width=0.45\linewidth]{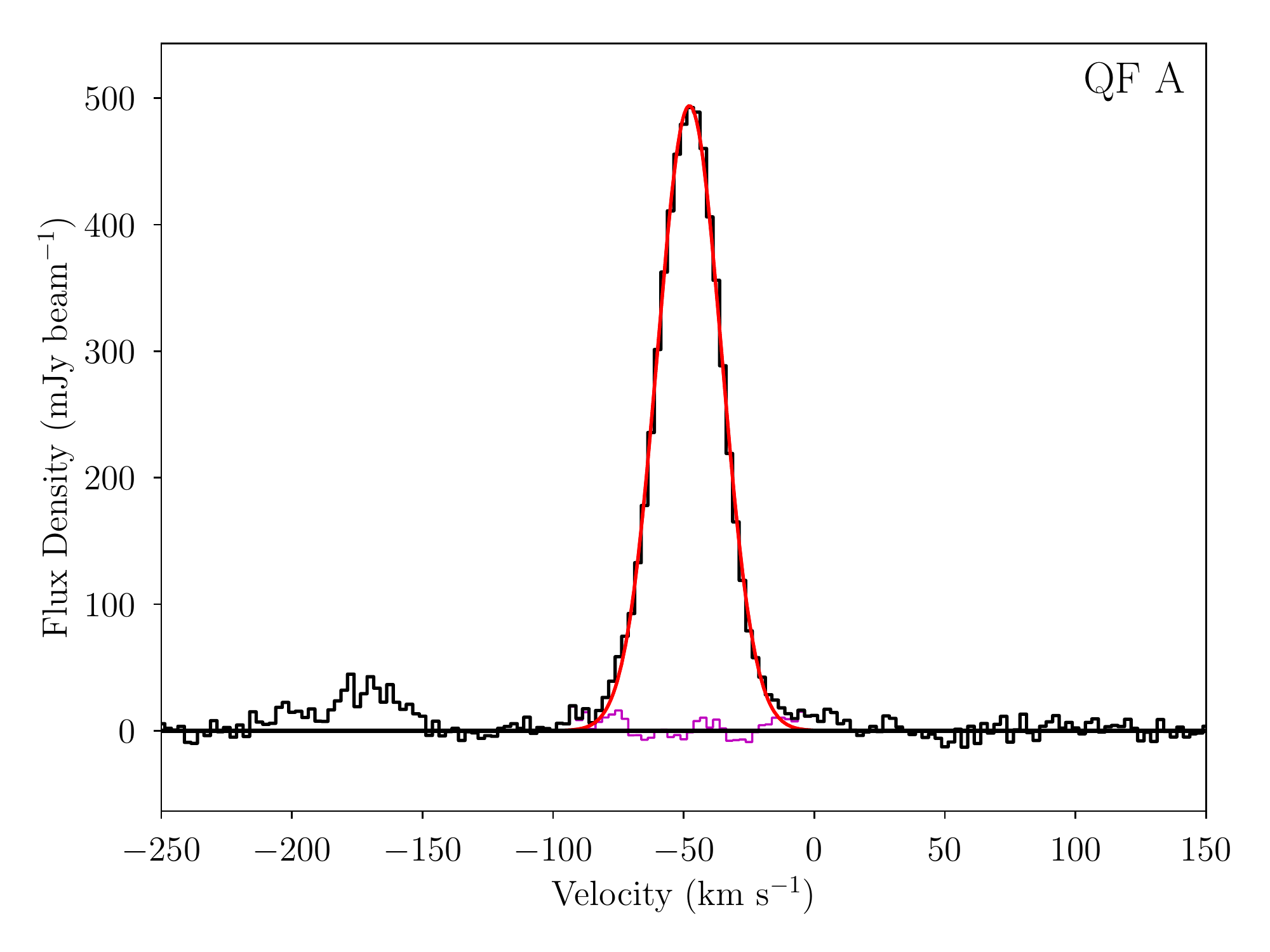}
  \includegraphics[width=0.45\linewidth]{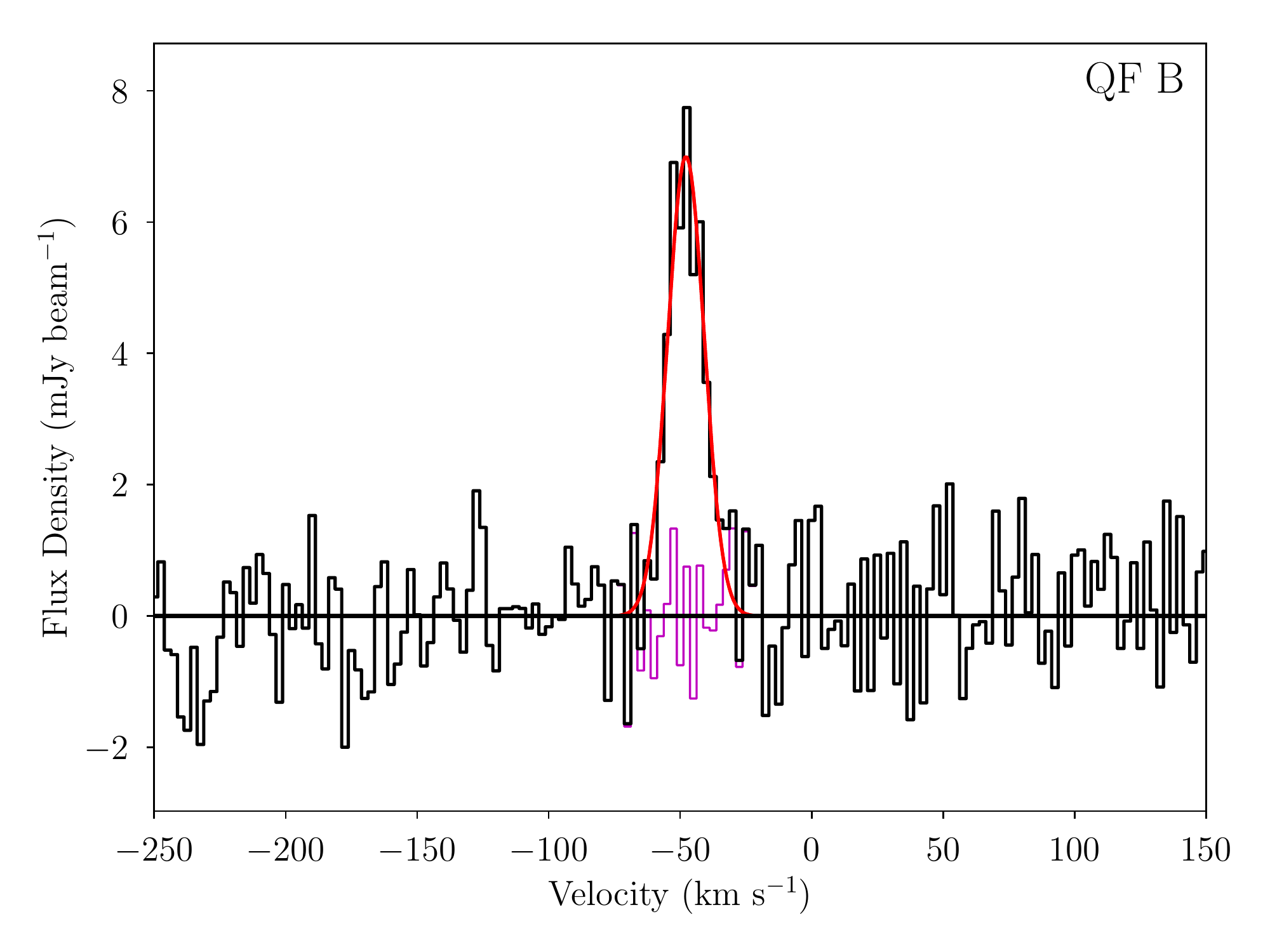} \\
  \includegraphics[width=0.45\linewidth]{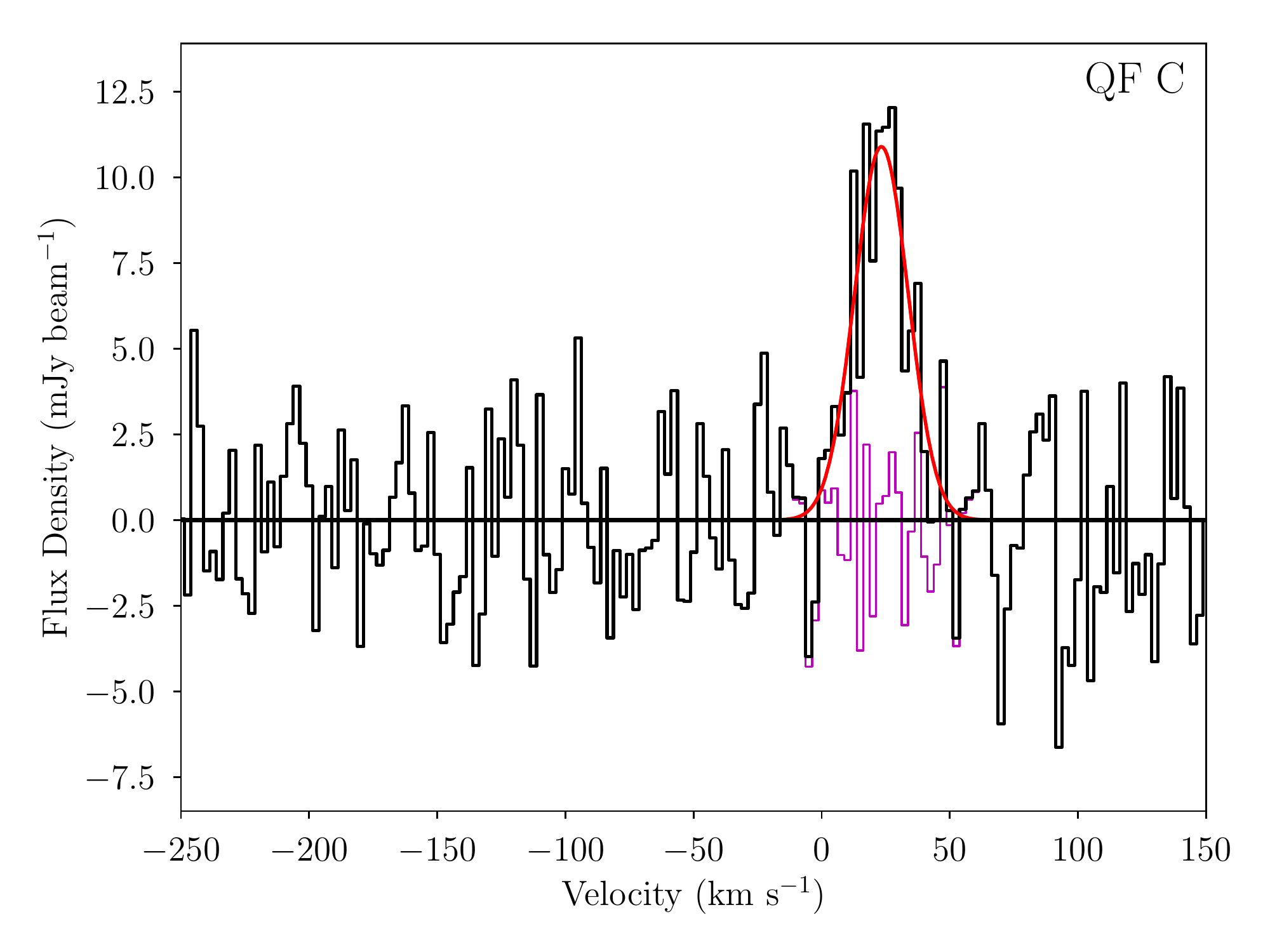}
  \includegraphics[width=0.45\linewidth]{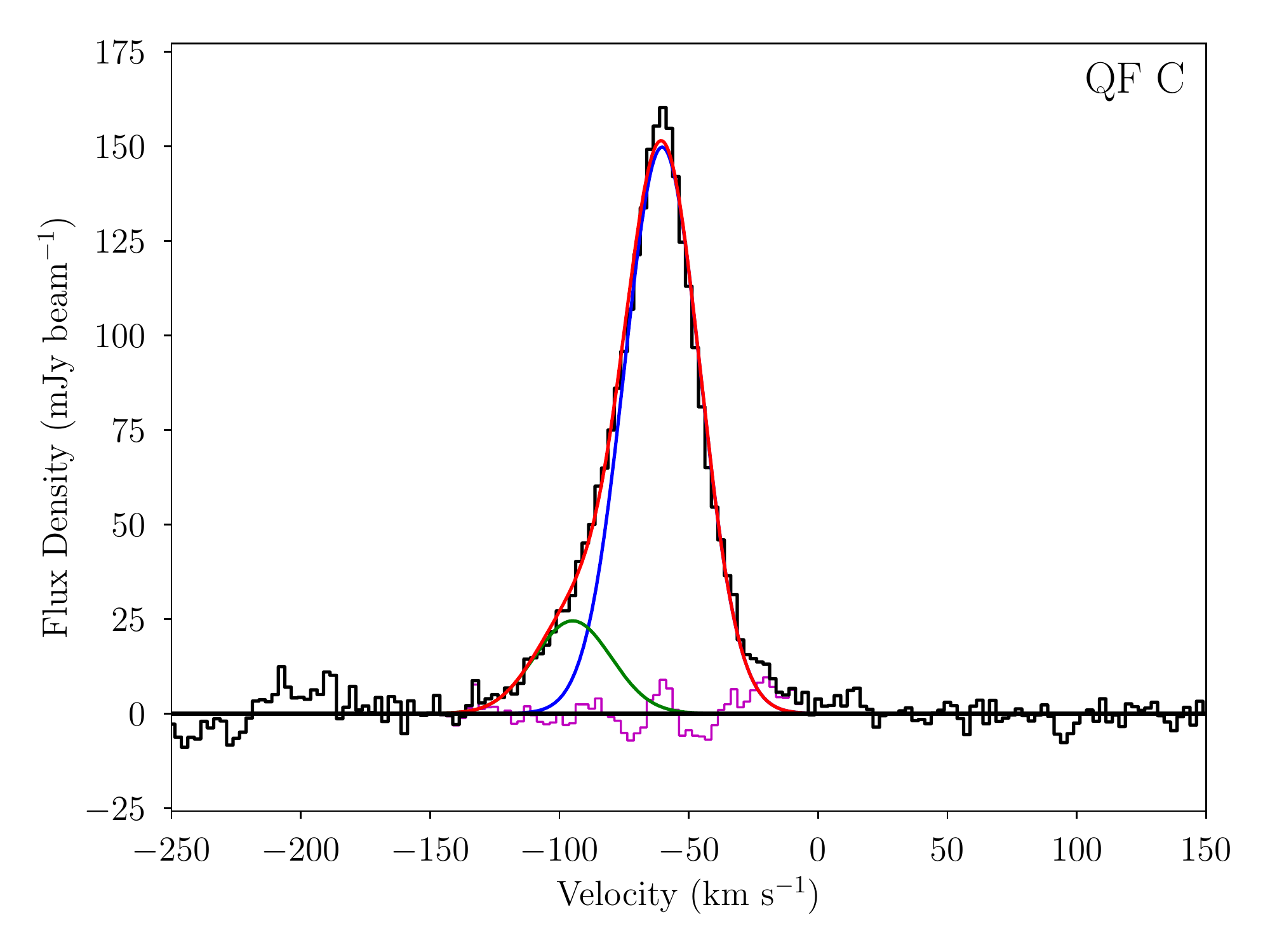}
  \caption{Representative \(<\)H\(n\alpha\)\(>\) spectra for each
    spectral quality factor (QF). The top-left panel is
    G333.129$-$00.439 and is a QF A spectrum (S/N\(>15\), un-blended),
    the top-right is G312.979$-$00.432 and is a QF B spectrum (\(15>\)
    S/N \(>10\)), the bottom-left panel is G311.866$-$00.238 and is a
    QF C spectrum (S/N \(<10\)), and the bottom-right is
    G332.823\(-\)00.550 and is a QF C spectrum (spectrally blended).
    The red curves are Gaussian fits to the data (or the sum of the
    two Gaussian components in G332.823\(-\)00.550), and the magenta
    curves are the fit residuals.}
  \label{fig:rrl_qf}
\end{figure*}

The calibration pipeline uses calibrator data to iteratively compute
calibration solutions, automatically flag bad data, apply calibration
solutions, and manually flag bad data. We typically repeat this
process two or three times for each observing session before the data
are clear of all obvious RFI (or otherwise bad data) and the
calibration solutions converge. The most common features we flag in
the SHRDS data are 1) persistent RFI missed by the automatic flagging
algorithms, 2) off-source antennas at the beginning of each scan, and
3) the first and last \({\sim}200\) channels of each spectral line
window. Spectral window 17 (8060\mhz, see Table~\ref{tab:backend}),
covering H93\(\alpha\), is nearly always flagged due to persistent,
broad-frequency RFI.

\subsection{Imaging}

Once the data from each observing session are fully calibrated, we
create a single data set for each field, from which we generate
images. The imaging part of our data reduction and analysis pipeline
is nearly fully-automated (see
Appendix~\ref{sec:app_data_reduction}). We first re-grid all of the
visibilities to a common kinematic local standard of rest (LSR or
LSRK\footnote{LSRK is defined by a solar motion of 20\kms\ in the
  direction (RA, Dec.) = (18h, +30\(^\circ\)) at epoch 1900.})
velocity frame with a channel width \(\Delta v_{\rm LSR} =
2.5\,\text{km s\(^{-1}\)}\). For each field, we then use the
\textit{CLEAN} algorithm to generate the following images and data
cubes: 1) a multi-scale, multi-frequency synthesis (MS-MFS) image
produced by combining the two 2\ghz\ continuum windows, 2) a MS-MFS
image of each 2\ghz\ continuum window, 3) a MS-MFS image of each
64\mhz\ bandwidth spectral line window, and 4) a multi-scale data cube
of each spectral line window.

The emission mechanisms of \hii\ regions allow us to take some
shortcuts in our imaging process. The thermal radio continuum emission
and RRL emission originate within the same volume of ionized gas. Thus
the morphology of the \hii\ region should not change from channel to
channel within a spectral line window. Therefore, to minimize
computation time, we use the \textit{CLEAN} masks from the MS-MFS
images of each spectral line window to \textit{CLEAN} that entire data
cube. We also do not perform any continuum subtraction.

For resolved sources, we can increase the surface brightness
sensitivity by \textit{uv-}tapering our data. By giving less weight to
the longer baselines when generating an image, \textit{uv}-tapering
increases the synthesized beam size and surface brightness
sensitivity.  Some fraction of our sources will be un-resolved and/or
in confused fields. In these cases \textit{uv}-tapering will worsen
our point-source sensitivity and spatial resolution. We therefore
generate two sets of images: one with no \textit{uv-}tapering and a
second with a \textit{uv-}taper to a synthesized HPBW of 100
arcseconds, approximately the synthesized HPBW of our lowest frequency
spectral line window. Depending on the morphology of the source, the
level of confusion within the field, and the scientific use, one of
these methods may be more useful than the other.

\subsection{Continuum Data Products}

The radio continuum image provides several measurable quantities for
each detected continuum source: position, peak continuum flux density,
total continuum flux density, and continuum spectral index. In this
first data release we only extract the positions and peak continuum
flux densities of our continuum detections associated with
\textit{WISE} Catalog \hii\ regions and \hii\ region candidates. In a
future data release, we will publish the total continuum flux
densities and continuum spectral indicies as well.

We use the 8 -- 10 \ghz\ MS-MFS continuum band image to identify
continuum sources within our primary beam. Although the full
4\ghz\ bandwidth continuum image has better sensitivity, the 8 -- 10
\ghz\ image has better spatial resolution. This higher-frequency image
will reveal more structure in confused fields. In each field, we
identify distinct continuum emission peaks by visual inspection. Each
distinct continuum peak is what we call a ``continuum source.'' In
this data release, we only identify continuum sources associated with
\textit{WISE} Catalog \hii\ regions or \hii\ region candidates. To be
a continuum source in our Bright Catalog, we require that the
continuum peak be within a circle centered on the \textit{WISE}
Catalog position with a radius equal to the infrared radius.

\subsection{RRL Data Products}

Our goal here is to create a catalog of \hii\ region positions and RRL
LSR velocities. We maximize our spectral sensitivity by averaging
every H\(n\alpha\) RRL transition and both polarizations to create a
stacked H\(n\alpha\) spectrum, denoted by \(<\)H\(n\alpha\)\(>\). Upon
the completion of the entire SHRDS, we will publish the spectra for
each detected individual RRL transition.

We extract and average spectra from our data in two ways, depending on
whether the data cube is \textit{uv}-tapered or not. For non-tapered
data cubes, we extract spectra from each of our spectral windows at
the location of the peak continuum source brightness. We remove all
poor-quality spectra, usually caused by un-flagged RFI or a very poor
baseline structure, then we use a weighted average to create the
\(<\)H\(n\alpha\)\(>\) spectrum. The weights for the \(i\)-th spectral
line window are given by \(w_i = S_{C,i}/{\rm rms}_i^2\), where
\(S_{C,i}\) is the continuum brightness and rms\(_i\) is the spectral
noise of the \(i\)-th spectral line window. Both the continuum
brightness and rms noise are estimated from the line-free regions of
the spectrum. For \textit{uv}-tapered data cubes, we first smooth each
of the spectral window cubes to a common beam size slightly larger
than the \textit{uv}-tapered beam size (typically 110
arcseconds). After removing spectral windows with poor-quality
spectra, we average the remaining cubes to create the
\(<\)H\(n\alpha\)\(>\) cube using the same weighting scheme as with
the non-tapered data. We extract the \(<\)H\(n\alpha\)\(>\) spectrum
at the location of the peak continuum source brightness.

For each \(<\)H\(n\alpha\)\(>\) spectrum, we identify the
line-free regions and use those to estimate the spectral rms noise and
to fit and subtract a third-order polynomial baseline. We then fit a
single Gaussian line profile to the spectrum to measure the RRL
brightness, full-width at half-maximum (FWHM) line width, LSR
velocity, and signal-to-noise ratio (S/N). We estimate S/N
following the \citet{lenz1992} method:
\begin{equation}
  {\rm S/N} = 0.7\left(\frac{S_L}{\rm rms}\right)\left(\frac{\Delta V}{\Delta v}\right)^{0.5} \label{eq:snr}
\end{equation}
where \(S_L\) is the peak line intensity, rms is the spectral noise,
\(\Delta V\) is the FWHM line width, and \(\Delta v\) is the channel
width (2.5\kms). In cases where multiple RRL components are visible,
we fit multiple Gaussian profiles.

\section{SHRDS: The Bright Catalog}

The SHRDS Bright Catalog contains the radio continuum and
\(<\)H\(n\alpha\)\(>\) RRL properties of the brightest \hii\ regions
in the survey. We observe 282 fields containing 149 previously known
\hii\ regions and 632 \hii\ region candidates in the Galactic
longitude range \(259^\circ < \ell < 344^\circ\). The majority of
these objects are too faint or large to be detected by the ATCA. They
are serendipitously observed because they are close to bright
\hii\ regions and \hii\ region candidates on the sky.  We detect at
least one radio continuum source in 275 fields (97.5\%) and at least
one \(<\)H\(n\alpha\)\(>\) RRL in 258 fields (91.5\%). We find radio
continuum and \(<\)H\(n\alpha\)\(>\) RRL emission towards 80 and 76
previously-known \hii\ regions, respectively, and towards 298 and 256
\hii\ region candidates, respectively. If instead we consider our
``nominal'' targets (\hii\ region candidates with predicted
6\ghz\ peak continuum flux densities brighter than \(60\,\text{mJy
  beam\(^{-1}\)}\) and predicted radio diameters less than 265
arcseconds, or known \hii\ regions with predicted radio diameters less
than 265 arcseconds), there are only 100 previously-known
\hii\ regions and 279 \hii\ region candidates in our fields. Of these
we detect radio continuum emission towards 72 (72\%) previously-known
\hii\ regions and 246 (88\%) \hii\ region candidates. We also find RRL
emission towards 76 (76\%) previously-known \hii\ regions and 230
(82\%) \hii\ region candidates. Many of these sources lie near the
edge of our primary beam which further decreases our detection rates.

\begin{figure*}
  \centering
  \includegraphics[width=0.45\linewidth]{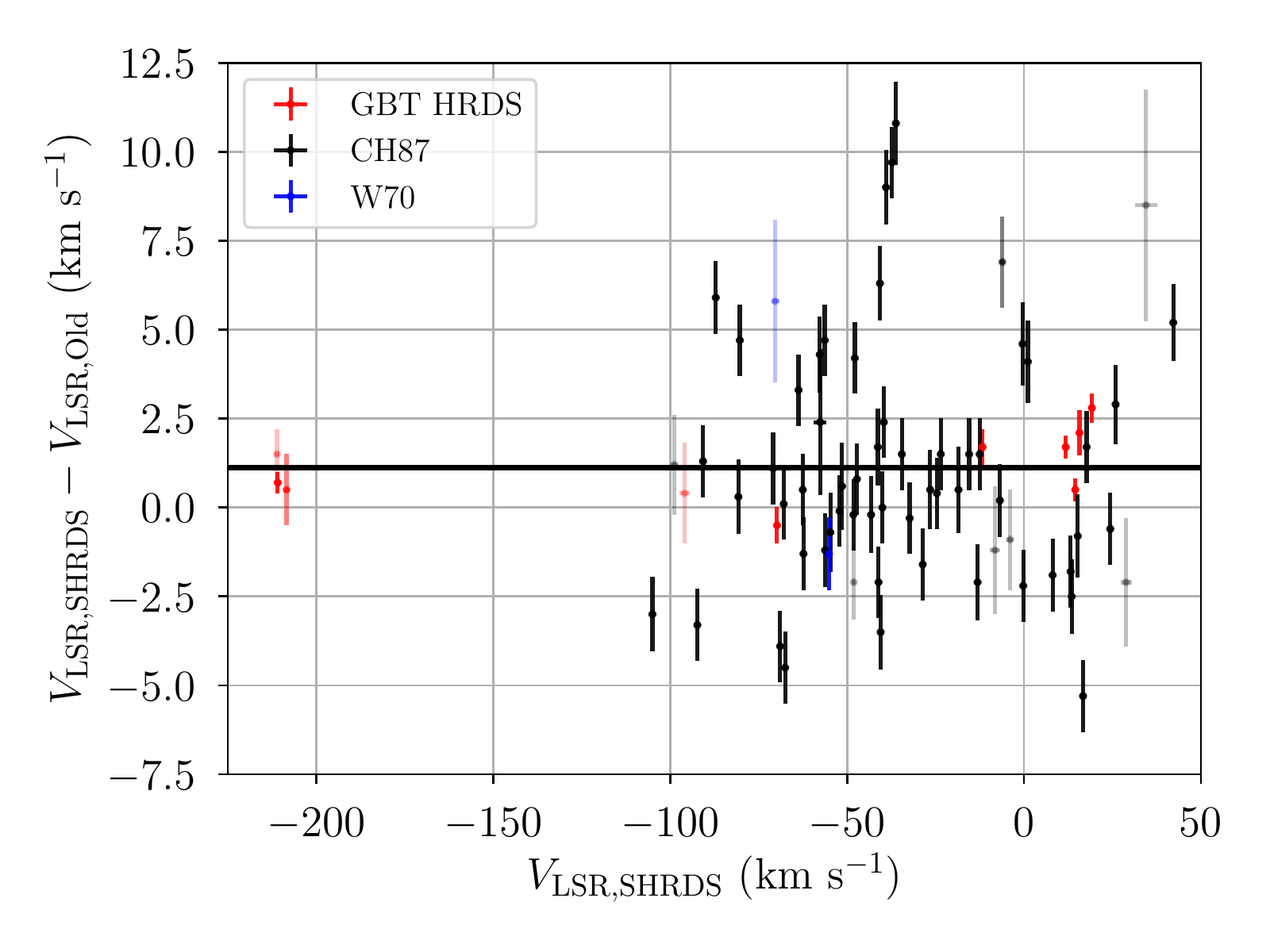}
  \includegraphics[width=0.45\linewidth]{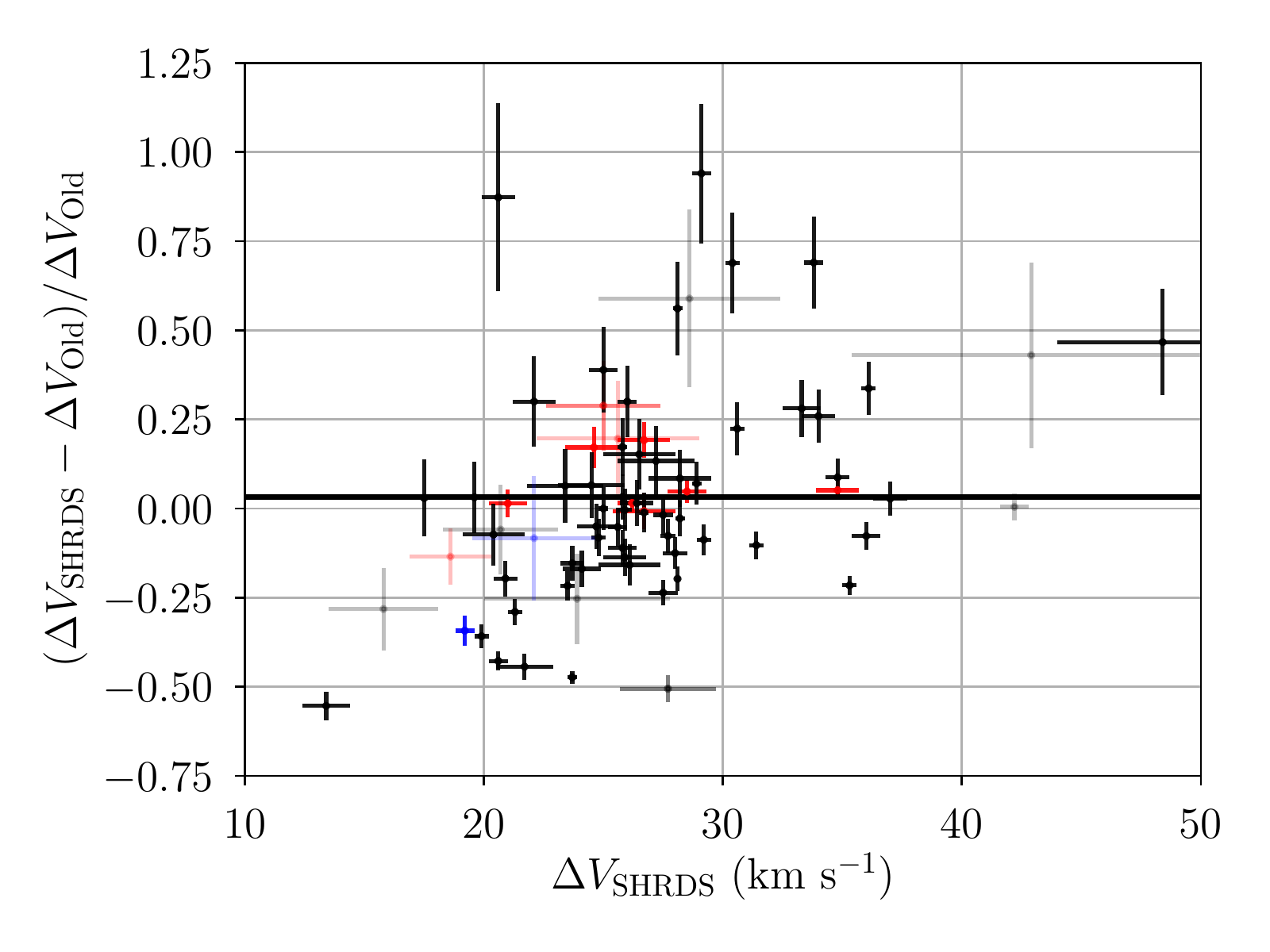} \\
  \includegraphics[width=0.45\linewidth]{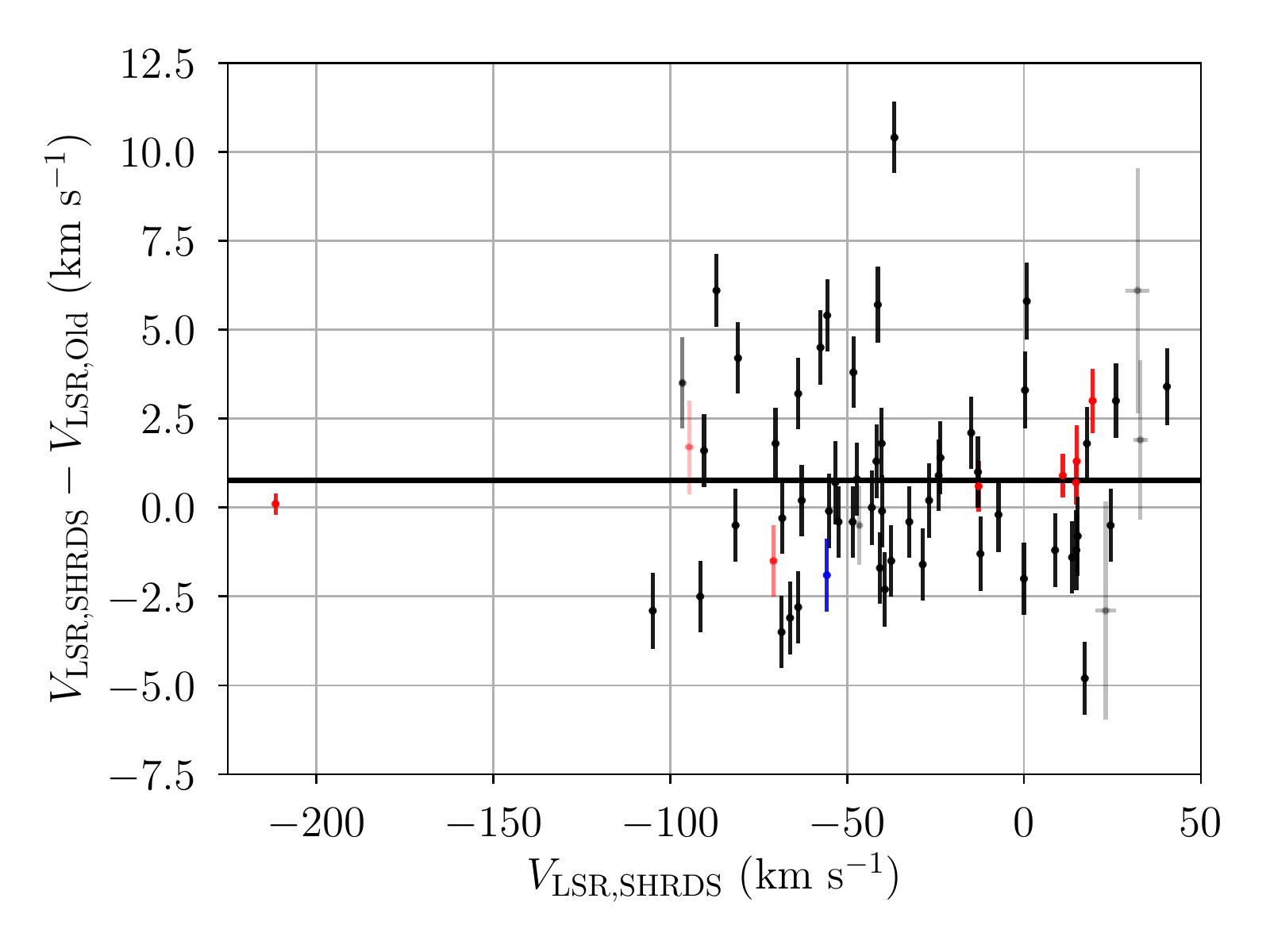}
  \includegraphics[width=0.45\linewidth]{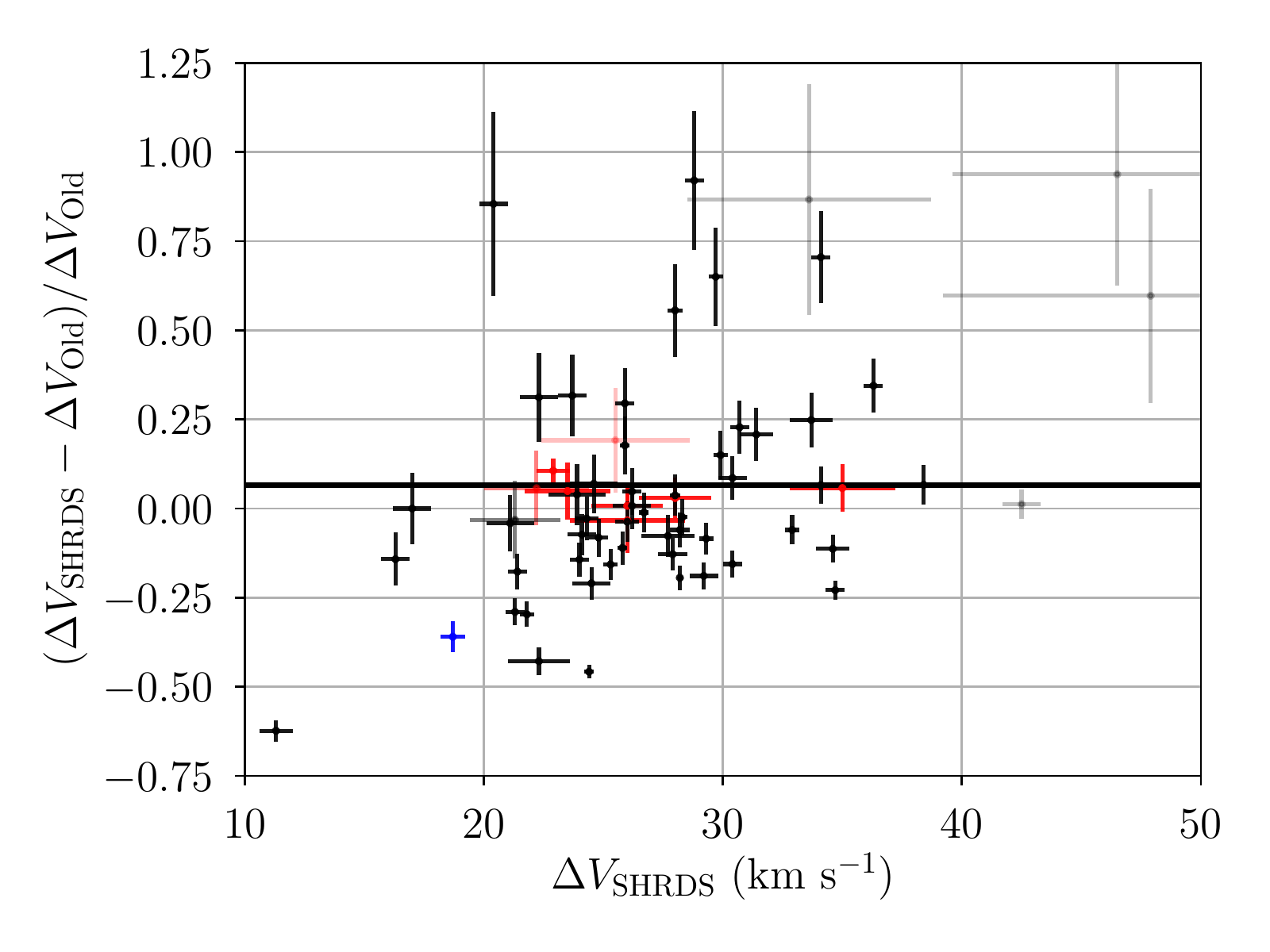}
  \caption{Differences between SHRDS and previously-measured
    \(<\)H\(n\alpha\)\(>\) LSR velocities, \(V_{\rm LSR}\) (left), and
    fractional differences between SHRDS and previously-measured FWHM
    line widths, \(\Delta V\) (right). The top two panels use the
    non-tapered data and the bottom two panels use the
    \textit{uv}-tapered data. Previous RRL measurements are from the
    GBT HRDS (black; \(n=10\) non-tapered; \(n=8\) tapered),
    \citet{caswell1987} (CH87; red; \(n=63\) non-tapered; \(n=56\)
    tapered), and \citet{wilson1970} (W70; blue; \(n=2\) non-tapered;
    \(n=1\) tapered). The transparency of the points represents the
    quality factor of the SHRDS RRL detection: QF A (not transparent),
    QF B (partially transparent), and QF C (mostly transparent). Six
    non-tapered SHRDS detections and five \textit{uv}-tapered
    detections with LSR velocities more than 15\kms\ different from
    the \citet{caswell1987} velocities are excluded. The mean of each
    sample is indicated by the horizontal solid line.}
  \label{fig:previous_compare}
\end{figure*}

In this catalog we also include data from the SHRDS pilot survey
\citep{brown2017}. We attempt to re-process these data using WISP with
mixed success. The fields observed in the SHRDS pilot survey did not
typically have adequate \textit{uv}-coverage to create high fidelity
images. Rather, \citet{brown2017} extracted spectra directly from the
\textit{uv} data and averaged them to create the
\(<\)H\(n\alpha\)\(>\) spectra. They report detections of
\(<\)H\(n\alpha\)\(>\) RRL emission from 7 previously-known
\hii\ regions and 36 out of 53 \hii\ region candidates. In our
re-processed pilot survey data, we find continuum emission from 7
previously-known \hii\ regions and 46 \hii\ region candidates, and
\(<\)H\(n\alpha\)\(>\) RRL emission from 7 previously-known
\hii\ regions and 26 \hii\ region candidates. Nearly all of our
detections are from the first epoch (2013) of the SHRDS pilot survey,
simply due to the better \textit{uv}-coverage of the first epoch
observations. Our re-processing discovers \(<\)H\(n\alpha\)\(>\) RRL
emission from 4 previously-known \hii\ regions and 1 \hii\ region
candidate \textit{not} in the pilot survey catalog. These sources are
not in the center of the field and thus were missed by the
\textit{uv}-spectra averaging method used by \citet{brown2017}. In the
following catalogs, we list our re-analyzed products, if available,
otherwise we give the values from \citet{brown2017}. Data taken
directly from \citet{brown2017} have an asterisk (*) appended to the
epoch. Three of the pilot survey ``previously-known'' \hii\ regions
(G290.323$-$02.984, G295.748$-$00.207, and G323.464+00.079) did not
have previous detections of H\(\alpha\) or RRL emission, so they are
listed as candidate \hii\ regions in the \textit{WISE} Catalog. These
sources are new SHRDS \hii\ region discoveries.

\subsection{Continuum Catalog}

Our continuum source detections are listed in
Table~\ref{tab:cont_notap} (non-tapered) and
Table~\ref{tab:cont_uvtap} (\textit{uv}-tapered). For each source, we
list the \textit{WISE} Catalog name, position of the peak radio
continuum emission, field containing the source, epoch, synthesized
beam area in the continuum image, separation between the observed
continuum peak position and infrared position, \(\Delta \theta\),
MFS-synthesized frequency of the continuum image, \(\nu_C\), peak
continuum flux density, \(S_C\), rms noise, rms\(_C\), and a ``quality
factor''. The peak continuum flux density and position are measured at
the location of the brightest pixel of the continuum source in the 4
\ghz\ bandwidth MS-MFS continuum image. The rms noise is estimated as
the rms across the entire residual image divided by the primary beam
response at the continuum source position. The quality factor, QF, is
a qualitative assessment of the accuracy of the peak continuum flux
density. QF A means the source is un-resolved, un-confused, and
located near the center of the primary beam, QF B means the source is
slightly resolved, slightly confused, and/or located off-center, and
QF C means the source is very resolved, very confused, and/or located
near the edge of the primary beam. The primary beam shape is not
modeled accurately by CASA beyond \({\sim}300\,\text{arcseconds}\)
from the field center, so continuum sources near the primary beam edge
may have flux densities in error by \(\gsim10\%\) (see
Appendix~\ref{sec:app_data_reduction}).  Figure~\ref{fig:cont_qf}
shows representative 8 -- 10 \ghz\ continuum band images for each QF:
G333.725+00.364 (QF A), G312.706+00.012 and G312.675+00.048 (QF B),
and G312.764+00.067 (QF C).

A given \textit{WISE} Catalog source may appear in multiple fields.
These cases, which we call ``multiple detections,'' nonetheless have
only a single entry in Tables~\ref{tab:cont_notap} and
\ref{tab:cont_uvtap}. The data for these targets have the highest
continuum QF and/or largest peak continuum flux-to-rms ratio. Every
field is an independent observation of the source, however, so we give
the continuum properties as measured in each field in
Appendix~\ref{sec:app_multiple_detections}. In the next data release
we will combine the data from each field to improve our sensitivity
and to measure the continuum properties of these sources more
accurately.

\begin{figure}
  \centering
  \includegraphics[width=\linewidth]{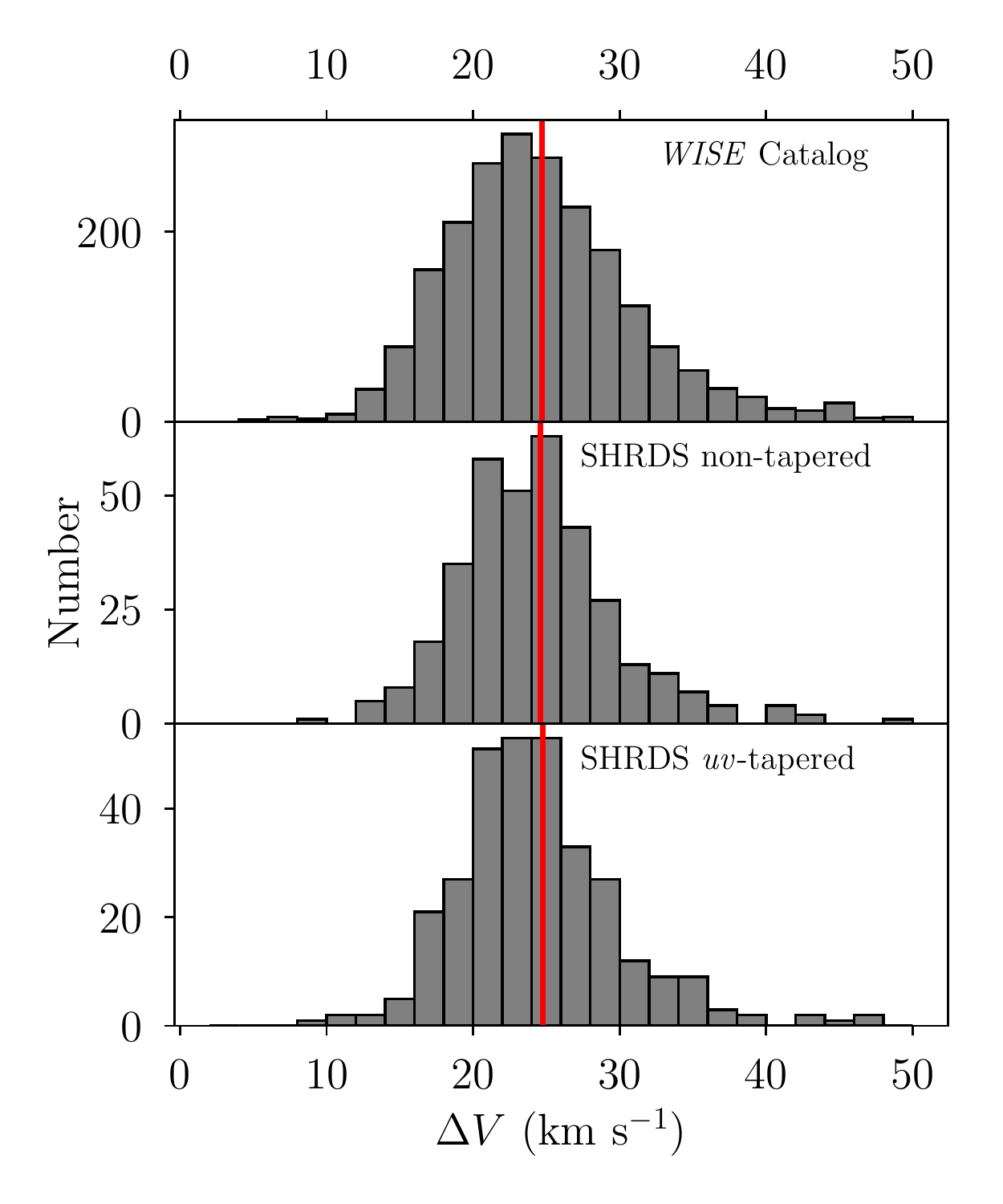}
  \caption{Distribution of \(<\)H\(n\alpha\)\(>\) RRL FWHM line
    widths, \(\Delta V\), for all previously-known \hii\ regions in
    the \textit{WISE} Catalog (top; \(n=2148\)), SHRDS non-tapered
    detections (middle; \(n=353\)), and SHRDS \textit{uv}-tapered
    detections (bottom; \(n=317\)), excluding 16 sources from the
    \textit{WISE} Catalog and two sources from the SHRDS with \(\Delta
    V > 50\,\text{km s\(^{-1}\)}\). The mean of each sample is
    indicated by the horizontal red line: 24.7\kms\ for the
    \textit{WISE} Catalog, 24.6\kms\ for the non-tapered SHRDS, and
    24.7\kms\ for the \textit{uv}-tapered SHRDS.}
  \label{fig:fwhm_histogram}
\end{figure}

\subsection{Radio Recombination Line Catalog}

We list the RRL emission properties for detected sources in
Table~\ref{tab:rrl_notap} (non-tapered \(<\)H\(n\alpha\)\(>\)) and
Table~\ref{tab:rrl_uvtap} (\textit{uv}-tapered
\(<\)H\(n\alpha\)\(>\)).  For each source, we give the \textit{WISE}
Catalog name, field, epoch, weighted-average frequency of the
\(<\)H\(n\alpha\)\(>\) RRL, \(\nu_L\), Gaussian fits to the peak line
intensity, \(S_L\), LSR velocity, \(V_{\rm LSR}\), and FWHM line
width, \(\Delta V\), rms spectral noise in the line-free region of the
spectrum, rms\(_L\), signal-to-noise ratio calculated using
Equation~\ref{eq:snr}, S/N, and a quality factor, QF. Like those
assigned in the continuum catalog, the QFs are qualitative assessments
of the accuracy of the Gaussian fits. Some of the
\(<\)H\(n\alpha\)\(>\) spectra show multiple RRL components. These
components may be distinct in velocity but they are often spectrally
blended. QF A means the RRL is un-blended with \(S/N > 15\). QF B
means the RRL is partially-blended but has distinct peaks and/or has
\(15 > S/N > 10\). QF C means the RRL is very blended with no distinct
peaks and/or has \(S/N < 10\). Finally, QF D is assigned to spectra
with no visible RRL. Figure~\ref{fig:rrl_qf} shows example
\(<\)H\(n\alpha\)\(>\) spectra for each QF: G333.129\(-\)00.439 (QF
A), G312.979\(-\)00.432 (QF B), and G311.866\(-\)00.238 and
G332.823\(-\)00.550 (QF C). Sources with multiple RRL components have
multiple rows in the tables, and the \textit{WISE} Catalog name is
appended by a letter, with ``a'' being the brightest RRL component,
``b'' being the second brightest, and so on.

We have multiple detections of RRLs for sources observed in separate
fields. Tables~\ref{tab:rrl_notap} and \ref{tab:rrl_uvtap} only
contain the \(<\)H\(n\alpha\)\(>\) Gaussian fits with the highest RRL
QF and/or S/N. The data for these sources as measured in each
independent field are listed in
Appendix~\ref{sec:app_multiple_detections}. In the next data release,
we will combine the \(<\)H\(n\alpha\)\(>\) spectra from each field to
increase our spectral sensitivity.

\section{Properties of Bright Catalog Nebulae}

For the longitude range \(259^\circ < \ell < 344^\circ\), the SHRDS
Bright Catalog increases the number of known Galactic \hii\ regions to
568 nebulae, an 82\% increase in the \hii\ region census of this
Galactic zone. With such a large sample of \hii\ regions in the third
and fourth Galactic quadrants, we can begin to get a global view of
star formation and Galactic structure across the entire disk.

\begin{figure*}
  \centering
  \includegraphics[width=\linewidth]{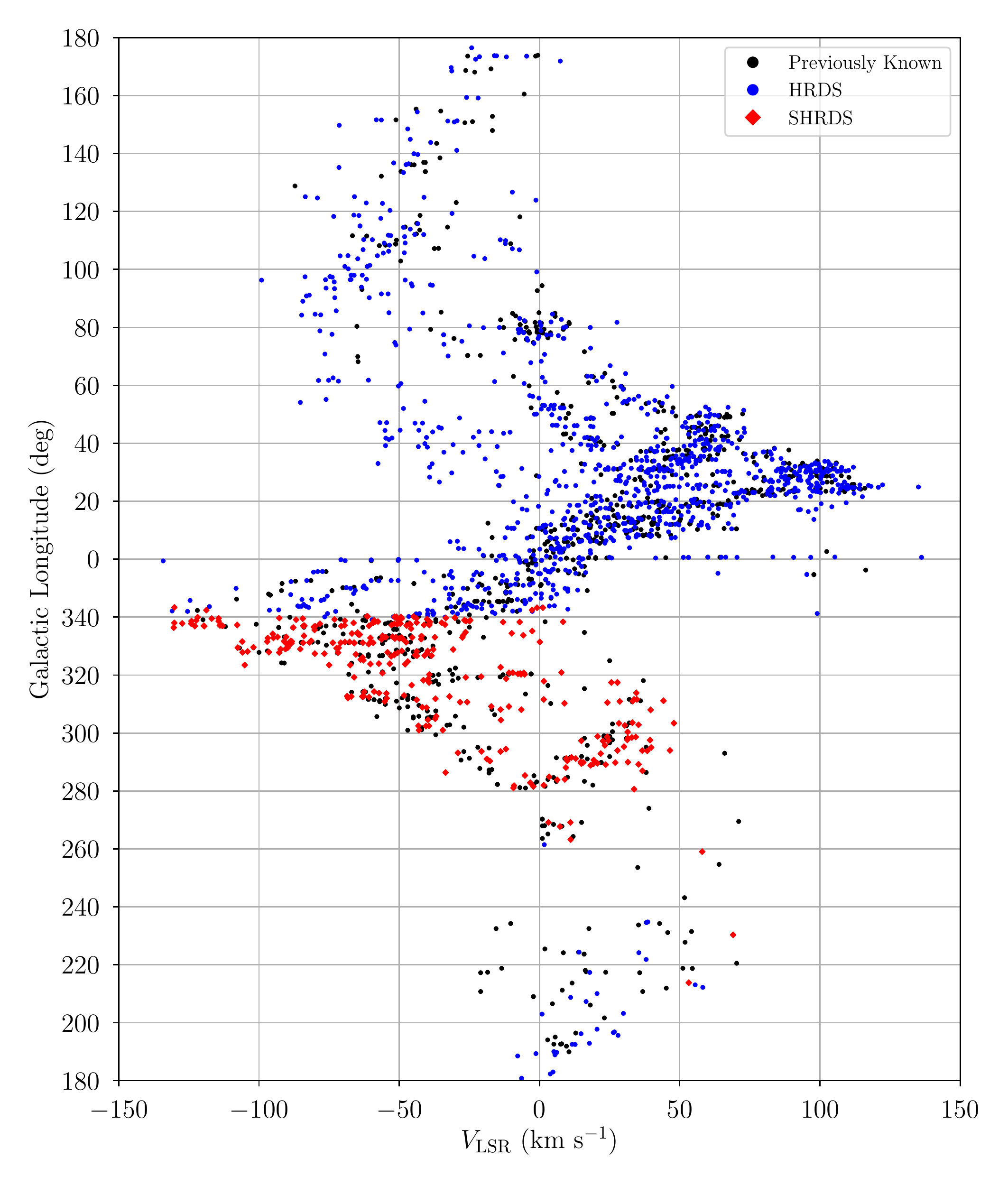}
  \caption{Galactic longitude, \(\ell\), as a function of LSR
    velocity, \(V_{\rm LSR}\), for all known Galactic \hii\ regions
    with \(|V_{\rm LSR}| < 150\kms\). Black points are \hii\ regions
    known prior to the GBT HRDS, blue points are \hii\ regions
    discovered by the GBT and Arecibo HRDS and their extensions, and
    red diamonds are \hii\ regions in the SHRDS Bright Catalog.}
  \label{fig:lv_full}
\end{figure*}

\begin{figure*}
  \centering
  \includegraphics[width=\linewidth]{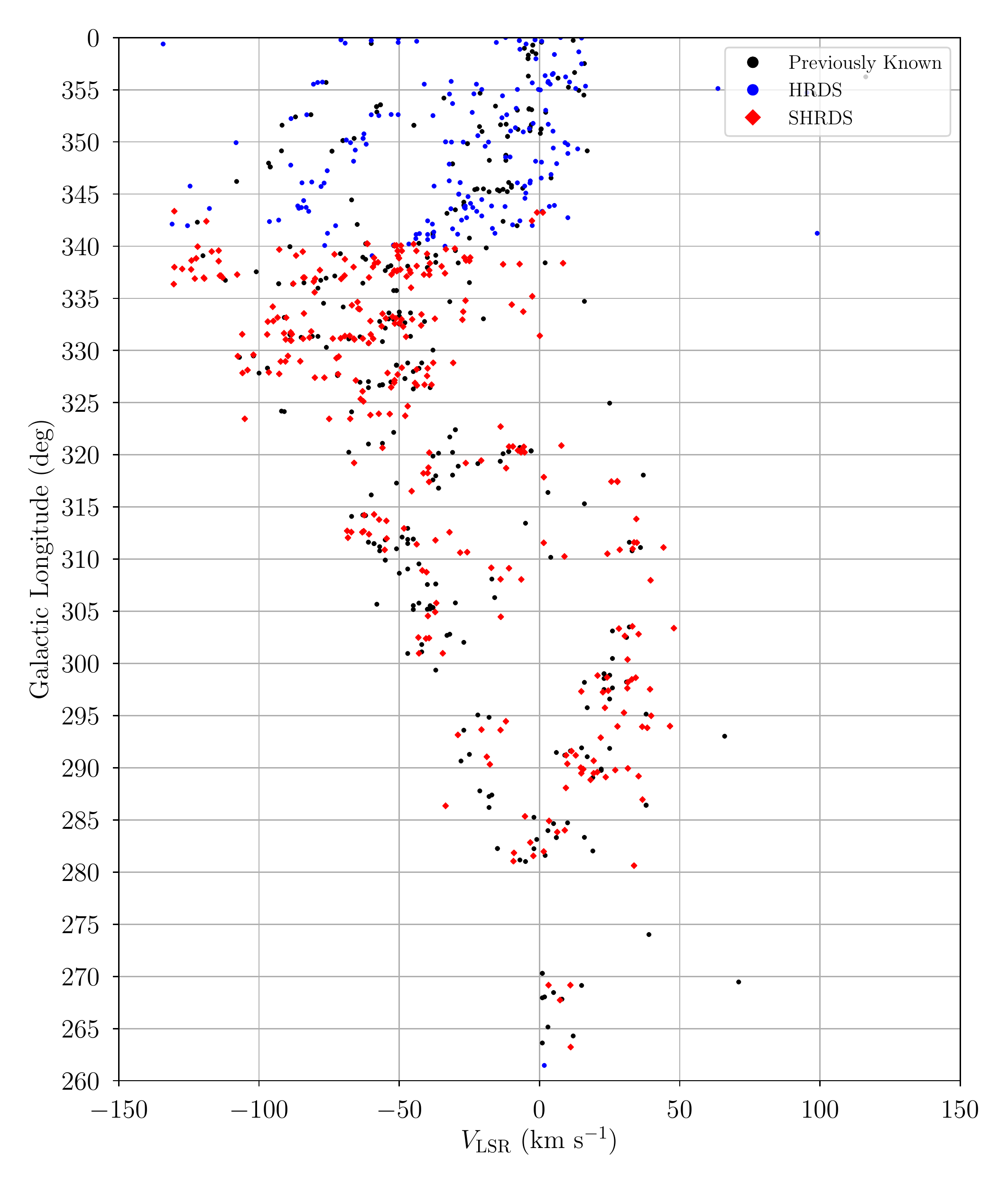}
  \caption{Same as Figure~\ref{fig:lv_full}, but zoomed in on the
    fourth Galactic quadrant.}
  \label{fig:lv_zoom}
\end{figure*}

To assess the validity of our data reduction and analysis pipeline, we
compare our measured RRL properties with those in the \textit{WISE}
Catalog for 76 previously-known \hii\ regions, 25 of which were
re-observed by the SHRDS for this
purpose. Figure~\ref{fig:previous_compare} shows the difference
between the SHRDS and previously-measured \(<\)H\(n\alpha\)\(>\) LSR
velocities and FWHM line widths as a function of the SHRDS-measured
values, for both the non-tapered and \textit{uv}-tapered SHRDS
data. These figures exclude six non-tapered and five
\textit{uv}-tapered SHRDS detections with LSR velocities more than
15\kms\ different from the \citet{caswell1987} velocities. These
sources are G289.806$-$01.242, G300.502-00.180 (non-tapered only),
G311.841$-$00.219, G311.866$-$00.238, G324.924$-$00.569, and
G333.681$-$00.441. Each of these nebulae are near an extended
\hii\ region or an \hii\ region complex with many other nebulae. We
suspect that the \citet{caswell1987} single dish survey is more
sensitive to the larger \hii\ regions and thus some LSR velocities are
incorrectly assigned to these smaller, nearby \hii\ regions in the
\textit{WISE} Catalog.

\begin{figure}
  \centering
  \includegraphics[width=\linewidth]{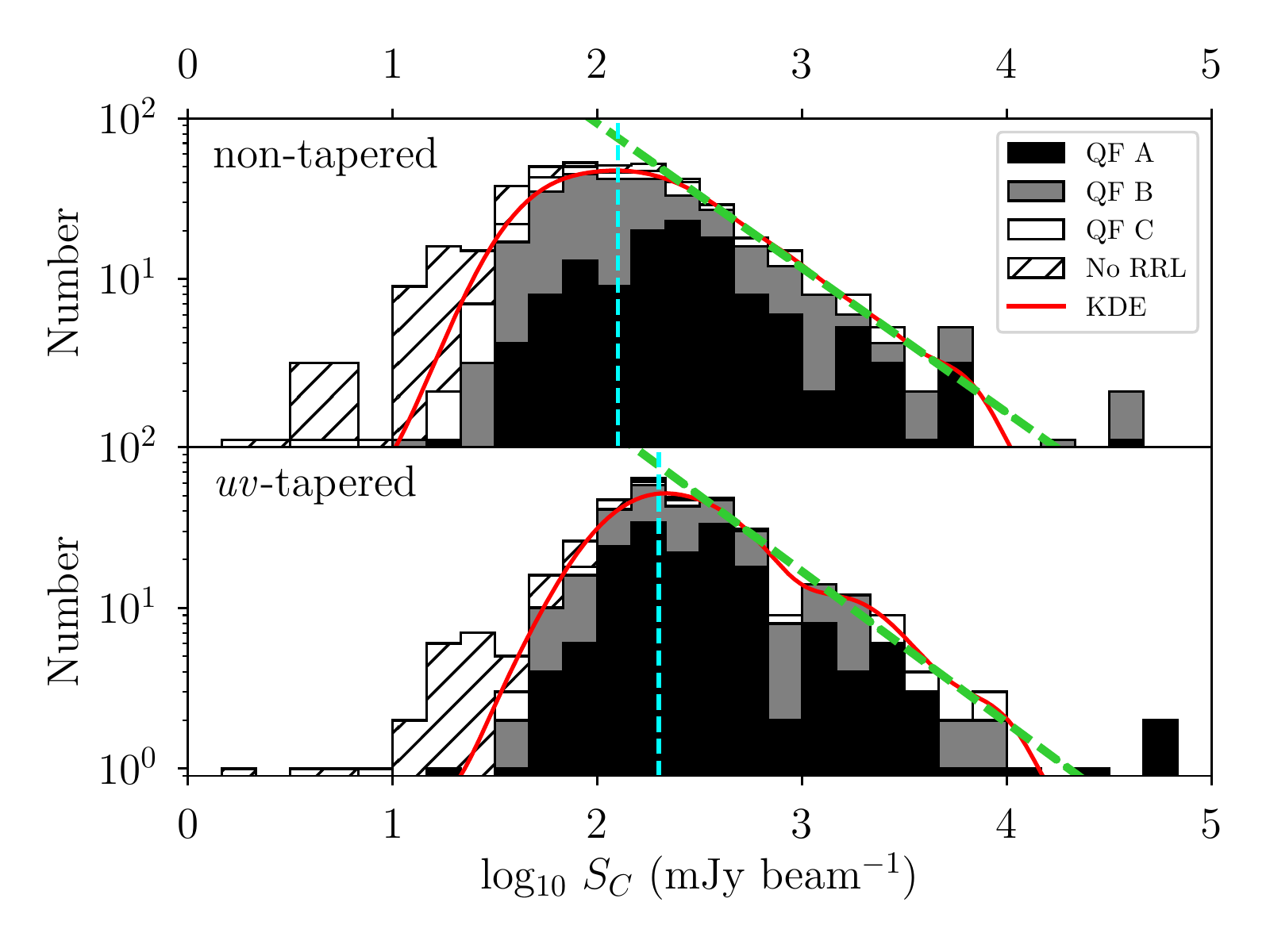}
  \caption{The distribution of peak continuum flux densities for the
    non-tapered (top) and \textit{uv}-tapered (bottom) SHRDS data. The
    shaded regions represent the contribution from sources of each
    continuum QF \textit{with a detected RRL} (QF A in black, QF B in
    gray, and QF C in white), and the hatched regions represent the
    contribution from sources \textit{without a detected RRL}. The red
    curves are the Gaussian kernel density estimator (KDE) fits to the
    distributions, and the green dashed curves are the power laws fit
    to a subset of the KDEs. The vertical cyan dashed lines represent
    the estimated continuum completeness limit of the SHRDS: 125 mJy
    beam\(^{-1}\) (non-tapered) and 200 mJy beam\(^{-1}\) (tapered) at
    \({\sim}7\ghz\).}
  \label{fig:completeness_cont}
\end{figure}

The SHRDS LSR velocities match previous measurements fairly well. The
LSR velocity differences are clustered around \(0\kms\) with slightly
more scatter towards larger SHRDS velocities. The mean velocity
difference is 1.1$\pm$0.4\kms with a standard deviation of 3.2
\kms\ for the non-tapered sample. For the \textit{uv}-tapered sample,
the mean difference is 0.8$\pm$0.3\kms with a standard deviation of
2.8\kms. This scatter is comparable to the spectral resolution of the
SHRDS and the \citet{caswell1987} survey (\({\sim}2.5\kms\)).

Similarly, the FWHM line width differences are centered around zero
with more scatter towards larger SHRDS line widths, some of which are
nearly a factor of two larger than previous measurements. We compute
the fractional difference between the SHRDS-measured FWHM line width,
\(\Delta\nu_{\rm SHRDS}\), and the previously-measured value,
\(\Delta\nu_{\rm Old}\), as \((\Delta\nu_{\rm SHRDS} - \Delta\nu_{\rm
  Old})/\Delta\nu_{\rm Old}\). The mean FWHM line width fractional
difference is 3$\pm$3\% with a standard deviation of 30\% for the
non-tapered sample, and is 7$\pm$8\% with a standard deviation of 32\%
for the \textit{uv}-tapered sample. The large scatter in the line
widths is probably an artifact of the different spatial scales probed
by an interferometer and a single dish telescope.

\begin{figure}
  \centering
  \includegraphics[width=\linewidth]{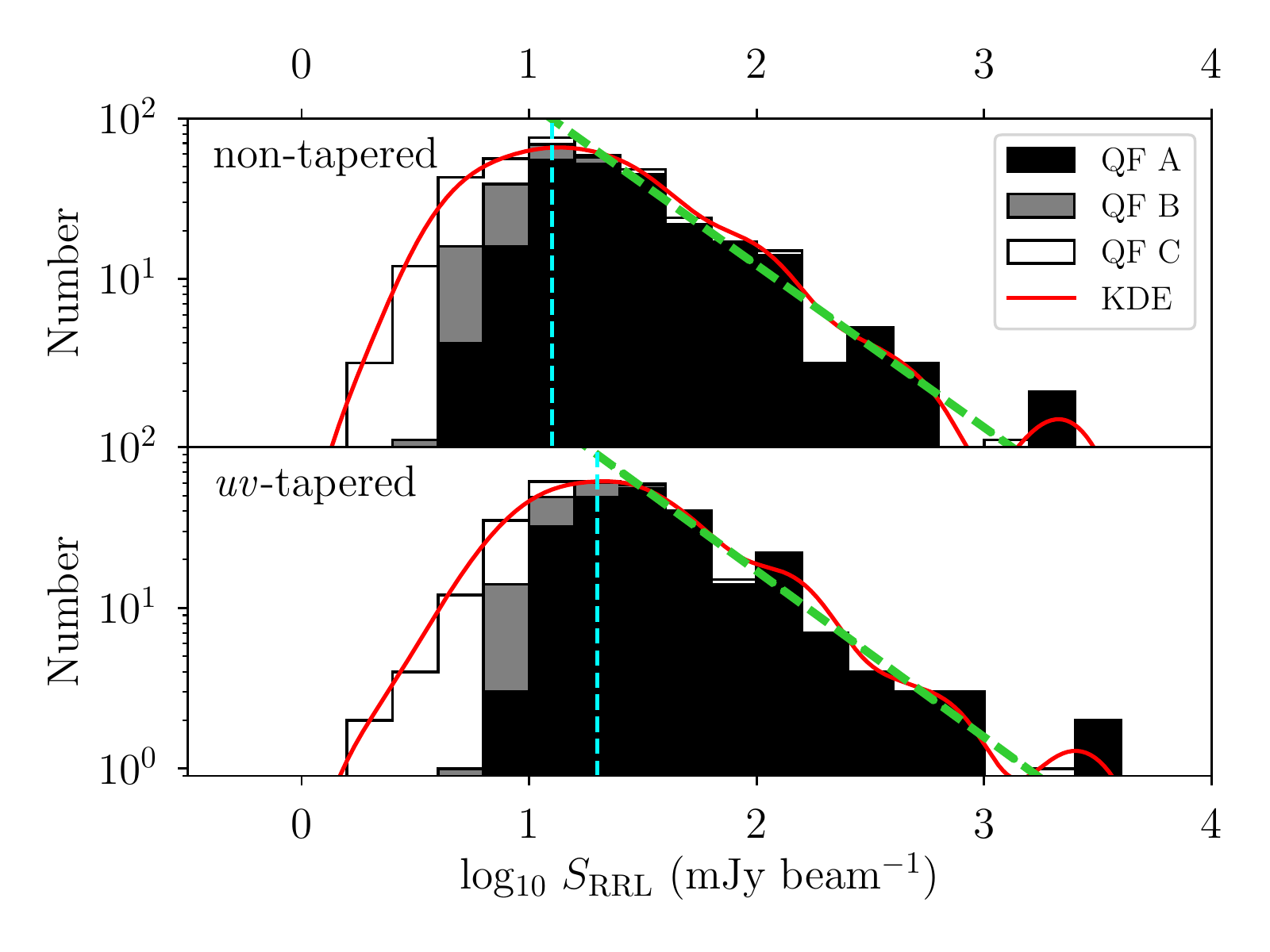}
  \caption{The distribution of \(<\)H\(n\alpha\)\(>\) RRL flux
    densities (see Figure~\ref{fig:completeness_cont}). We estimate
    the \({\sim}7\ghz\) RRL completeness to be 12.5 mJy beam\(^{-1}\)
    in the non-tapered data and 20 mJy beam\(^{-1}\) in the
    \textit{uv}-tapered data.}
  \label{fig:completeness_line}
\end{figure}

There is no difference in the distribution of \(<\)H\(n\alpha\)\(>\)
RRL FWHM line widths in the SHRDS and those in the \textit{WISE}
Catalog, nor a difference between the non-tapered and
\textit{uv}-tapered FWHM line width
distributions. Figure~\ref{fig:fwhm_histogram} shows the RRL line
widths of all known \hii\ regions with RRL measurements in the
\textit{WISE} Catalog and all of the RRL line widths measured in the
SHRDS, both non-tapered and \textit{uv}-tapered. Each distribution
peaks near \({\sim}23\,\text{km s\(^{-1}\)}\) and has a long tail
towards large FWHMs. The mean FWHM line widths are \(25\,\text{km
  s\(^{-1}\)}\) in each sample. The similarity of the SHRDS and
\textit{WISE} Catalog line width distributions implies that the
physical conditions (e.g., internal turbulence) of \hii\ regions are
comparable between the SHRDS survey zone and the rest of the Galaxy.

The Galactic longitude-velocity (\(\ell\)-\(v\)) diagram that includes
the nebulae discovered here shows interesting structure in the fourth
Galactic quadrant. Figure~\ref{fig:lv_full} is the \(\ell\)-\(v\)
diagram of all previously-known and SHRDS-discovered \hii\ regions
with LSR velocities \(|V_{\rm LSR}| < 150\,\text{km s\(^{-1}\)}\), and
Figure~\ref{fig:lv_zoom} is zoomed in on the fourth Galactic
quadrant. The SHRDS nebulae are constrained to \(V_{\rm LSR} <
50\kms\) in the fourth quadrant, likely due to a lack of
sensitivity. In the next data release we will target fainter and,
perhaps, more distant \hii\ regions, which may fill in this region of
the \(\ell\)-\(v\) diagram.

\subsection{Survey Completeness}

Following the strategy of \citet{anderson2011}, we estimate the
completeness of the Bright Catalog to be the point at which the flux
density distribution of SHRDS nebuale begins to deviate from a power
law. Assuming that the Galactic \hii\ region luminosity function is a
power law \citep[e.g.][]{smith1989,mckee1997} and that the
distribution of nebuale is relatively smooth across the disk, the
distribution of \hii\ region flux densities should be a power law as
well. In this data release, we only measure peak flux densities.
These flux densities will underestimate the total flux densities of
resolved sources, and therefore this analysis is not a representation
of the true \hii\ region luminosity function. In the next data release
we will measure the total flux densities of all SHRDS nebulae and
reassess the SHRDS completeness limit.

Figure~\ref{fig:completeness_cont} shows the distribution of
SHRDS peak continuum flux densities in both the non-tapered and
\textit{uv}-tapered data. Both distributions follow a power law at the
brighter end.  We estimate the completeness of the continuum catalog
as the point when these distributions begin to deviate from a power
law: \({\sim}125\,\text{mJy beam\(^{-1}\)}\) in the non-tapered
catalog and \({\sim}200\,\text{mJy beam\(^{-1}\)}\) in the
\textit{uv}-tapered catalog. Similar distributions are shown in
Figure~\ref{fig:completeness_line} for the distribution of
\(<\)H\(n\alpha\)\(>\) RRL flux densities. The completeness of the RRL
catalog as inferred from these distributions is ten times lower than
that of the continuum catalog, as expected for a typical
RRL-to-continuum intensity ratio of 0.01 at these frequencies.

\begin{figure}
  \centering
  \includegraphics[width=\linewidth]{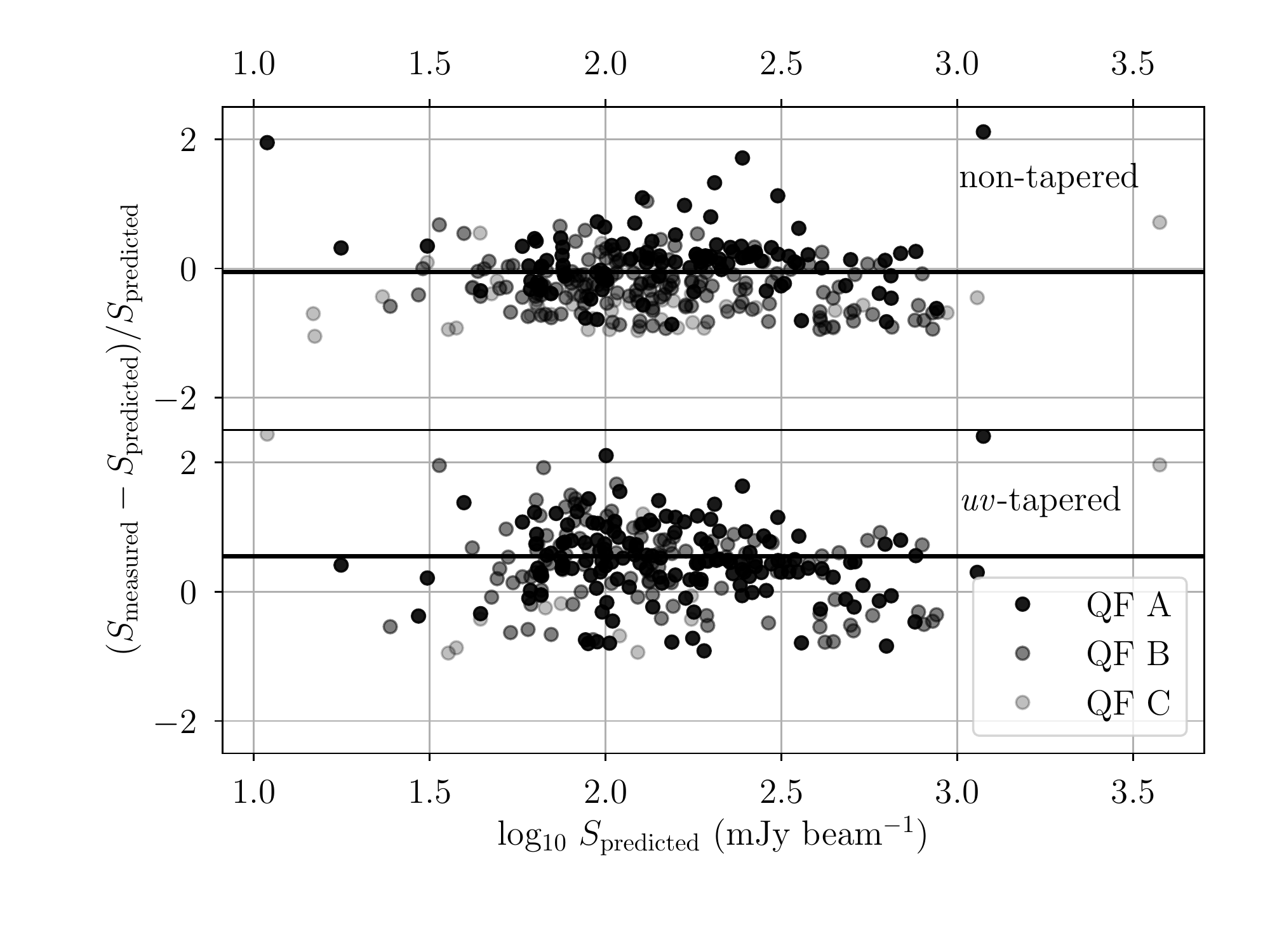}
  \caption{The fractional difference between the measured 6\ghz\ peak
    continuum flux densities and predicted flux densities as
    extrapolated from SUMSS. There are 313 non-tapered continuum
    detections (top) and 278 \textit{uv}-tapered continuum detections
    (bottom) with SUMSS predicted fluxes in the \textit{WISE}
    Catalog. Five sources with differences \(>250\%\) are excluded
    from each sample for clarity. The SHRDS peak continuum flux
    densities are extrapolated from their measured frequencies to
    6\ghz\ assuming a spectral index of \(\alpha = -0.1\). The
    transparency of the points represents the continuum quality factor
    of the detection: QF A (opaque), QF B (slightly transparent), QF C
    (mostly transparent). The horizontal solid lines indicate the mean
    fractional difference, -6\% (non-tapered) and 54\%
    (\textit{uv}-tapered).}
  \label{fig:sumss_scatter}
\end{figure}

Our non-tapered completeness limit (\({\sim}125\,\text{mJy
  beam\(^{-1}\)}\)) is more than twice our target selection criterion
(\(60\,\text{mJy beam\(^{-1}\)}\)). This difference is likely due to
the uncertainty in extrapolating \hii\ region flux densities from
SUMSS 843\mhz\ to 6\ghz.  Figure~\ref{fig:sumss_scatter} shows the
fractional difference between the measured and predicted peak
continuum flux densities for both our non-tapered and
\textit{uv}-tapered catalogs.  Here the SHRDS continuum fluxes are
scaled from their observed frequencies to 6\ghz\ assuming a spectral
index \(\alpha = -0.1\). The mean difference for the non-tapered data
is \(-6\%\) (over-predicted) and for the \textit{uv}-tapered data is
\(54\%\) (under-predicted). In both cases the standard deviation is
\({\sim}100\%\). \citet{anderson2011} found a similar scatter when
extrapolating from \({\sim}1.4\ghz\) to \(10\ghz\). This suggests that
many \hii\ regions are optically thick at and below 1.4\ghz.

\section{Summary and Future Work}

The SHRDS has already nearly doubled the number of known Galactic
\hii\ regions in the longitude range \(259^\circ < \ell <
344^\circ\). In this first data release, the Bright Catalog, we report
the detection of 256 new nebulae. We observe 282 fields and find
continuum emission towards \hii\ regions or \hii\ region candidates in
275 (97.5\%) and RRL emission in 258 (91.5\%) of them. We estimate
that the SHRDS Bright Catalog is complete for all \hii\ regions with
7\ghz\ peak continuum flux densities brighter than \(125\,\text{mJy
  beam\(^{-1}\)}\) in the surveyed zone.

We detect RRL emission from 76 previously-known \hii\ regions and find
that the SHRDS RRL properties are similar to previous
measurements. The mean LSR velocity difference between our
measurements and those in the \textit{WISE} Catalog is \(1.1\,\text{km
  s\(^{-1}\)}\) with a standard deviation of \(3.2\kms\). This scatter
is likely due to the velocity resolution of the SHRDS and previous
observations. The mean FWHM line width difference is \(3\%\) with a
standard deviation of \(30\%\). All previous \hii\ region surveys used
single dish telescopes, thus the scatter in FWHM line width
differences is likely to be a consequence of the different spatial
scales probed by the ATCA.

The distribution of SHRDS line widths is nearly identical to that of
all previously-known \hii\ regions in the \textit{WISE} Catalog. The
physical conditions, such as internal turbulence, of SHRDS
\hii\ regions are thus similar to those of the Galactic population of
\hii\ regions as a whole.

In this data release we provide only the positions, continuum flux
densities, and stacked \(<\)H\(n\alpha\)\(>\) RRL properties for our
sample of bright \hii\ regions. Much more insight will be gained by
looking at the intermediate data products, such as the continuum
spectral energy distributions, RRL spectral energy distributions,
RRL-to-continuum intensity ratio distributions, etc. The next SHRDS
data release will add \(\sim\)200 more \hii\ regions to the SHRDS
catalog, bringing the total number of newly-discovered nebulae to
\(\sim\)500. We will also publish the intermediate data products that
will allow for a more detailed analysis of individual \hii\ regions.

A complete catalog of Galactic \hii\ regions in the third and fourth
Galactic quadrants will drastically improve our understanding of
Galactic high-mass star formation, spiral structure, and metallicity
structure. With \hi\ emission/absorption observations toward these
newly-discovered \hii\ regions, we will resolve the kinematic distance
ambiguity of a subset of our sample and create the most complete
face-on map of Galactic \hii\ regions to date.

\acknowledgments

T.V.W. is supported by the NSF through the Grote Reber Fellowship
Program administered by Associated Universities, Inc./National Radio
Astronomy Observatory, the D.N. Batten Foundation Fellowship from the
Jefferson Scholars Foundation, the Mars Foundation Fellowship from the
Achievement Rewards for College Scientists Foundation, and the
Virginia Space Grant Consortium. L.D.A. is supported in part by NSF
grant AST-1516021. T.M.B. is supported in part by NSF grant
AST-1714688. J.R.D. is the recipient of an Australian Research Council
(ARC) DECRA Fellowship (project number DE170101086).

The Green Bank Observatory and National Radio Astronomy
Observatory are facilities of the National Science Foundation operated
under cooperative agreement by Associated Universities, Inc.

The Australia Telescope Compact Array is part of the Australia
Telescope National Facility, which is funded by the Australian
Government for operation as a National Facility managed by CSIRO.

We thank the anonymous referee for their valuable comments and
suggestions, which improved the quality of this paper.

\facility{ATCA}

\software{Astropy \citep{astropy2013},
  CASA \citep{mcmullin2007},
  Matplotlib \citep{matplotlib2007},
  NumPy \& SciPy \citep{numpyscipy2011},
  Python (\url{https://www.python.org/}),
  WISP \citep{wisp}}

\bibliography{manuscript}

\appendix
\section{WISP: A General Radio Interferometry Data Reduction Package}\label{sec:app_data_reduction}

Here we introduce the data reduction and analysis tools developed for
the SHRDS: the Wenger Interferometry Software Package
\citep[WISP;][]{wisp}\footnote{\url{https://doi.org/10.5281/zenodo.2225273}}. WISP
is a collection of \textit{Python} code implemented through the
\textit{Common Astronomy Software Applications} package
\citep[CASA;][]{mcmullin2007}. Its generic and modular framework is
designed to handle any continuum or spectral line radio interferometry
data. We are motivated to create our own data reduction pipeline for
three reasons: 1) the large quantity of data in the SHRDS, 2) missing
functionality in CASA, and 3) the lack of automatically-generated
quality diagnostics in existing pipelines.

Our primary motivation for WISP is the overwhelmingly large quantity
of data produced in the SHRDS. The calibration process for a single
\({\sim}8\) hour observing session of the SHRDS, for example, takes
\({\sim}16\) hours of raw computing time on a reasonably powerful
machine, not including the downtime between individual calibration
tasks, while inspecting the data, etc. Including this overhead, the
entire calibration process for a single observing session can take a
week if done interactively.  This is impractical considering the
hundreds of hours of data accumulated by the SHRDS.  With WISP, all of
the calibration steps are automated so that there is no downtime
between the completion of one task and the start of the next. The
total time it takes to calibrate a single observing session using WISP
is reduced to only \({\sim}24\) hours including the manual data
inspection.

The imaging process is equally as time consuming if done
interactively. To \textit{CLEAN} a single spectral line window data
cube for a single field can take \({\sim}15\) minutes. Given the
\({\sim}300\) fields and \({\sim}20\) spectral line windows, the total
computation time required to generate all of the images for the SHRDS
Bright Catalog is nearly two months. Interactive cleaning
(i.e. determining \textit{CLEAN} masks by hand) would add
substantially to the imaging time, so WISP uses automatic
\textit{CLEAN} mask generation and other tricks to automate the entire
imaging process.

The SHRDS Bright Catalog data grows from about 10 Terabytes in its raw
form to nearly 50 Terabytes after each field and spectral window are
imaged (``data reduction'' should really be called ``data expansion''
when discussing radio interferometry). All of the images and data
cubes generated for a single source total about 150 Gigabytes. WISP
(via CASA) generates several intermediate images and data cubes for
each field and spectral window, most of which are not needed for the
final data analysis. By converting the required images and cubes to
Flexible Image Transport System (FITS) files and deleting all
CASA-generated images and cubes, we reduce the disk space for a single
source to about 30 Gigabytes.

Our second motivation for WISP is a lack of some required
functionality in CASA. Most importantly for the SHRDS is the inability
to interpolate through missing or flagged data.  The ATCA correlator,
CABB, has several known bad channels (``birdies'').  If these channels
are flagged throughout \textit{all} of the data, CASA fails to
correctly account for the missing data when re-gridding the velocity
axis of our spectral line windows. This failure produces large,
periodic ripples in the spectra. WISP has the functionality to
seamlessly interpolate through any known bad channels by directly
editing the visibility data. Without missing data the velocity axis
re-gridding works correctly.

Finally, existing pipelines (at least those that existed when the
SHRDS began) lack the ability to automatically generate calibration
and imaging diagnostics to assess the quality of the data reduction
process. After the calibration process, for example, we wish to
inspect the derived calibration solutions and the calibrated
visibilities. In CASA, this requires the user to run a task, wait
(sometimes several minutes) for the plot(s) to be generated, inspect
the plots, then repeat for the next calibration solution or visibility
set. In total this can take several hours depending on the number of
calibrator sources and calibration solutions to inspect. WISP
automatically generates all of the necessary calibration solution and
visibility plots and compiles them into a single document. We can then
quickly scroll through the plots and identify any issues.

Below we describe in detail the specific calibration and imaging steps
performed by WISP.

\begin{figure}
  \centering
  \includegraphics[width=\linewidth]{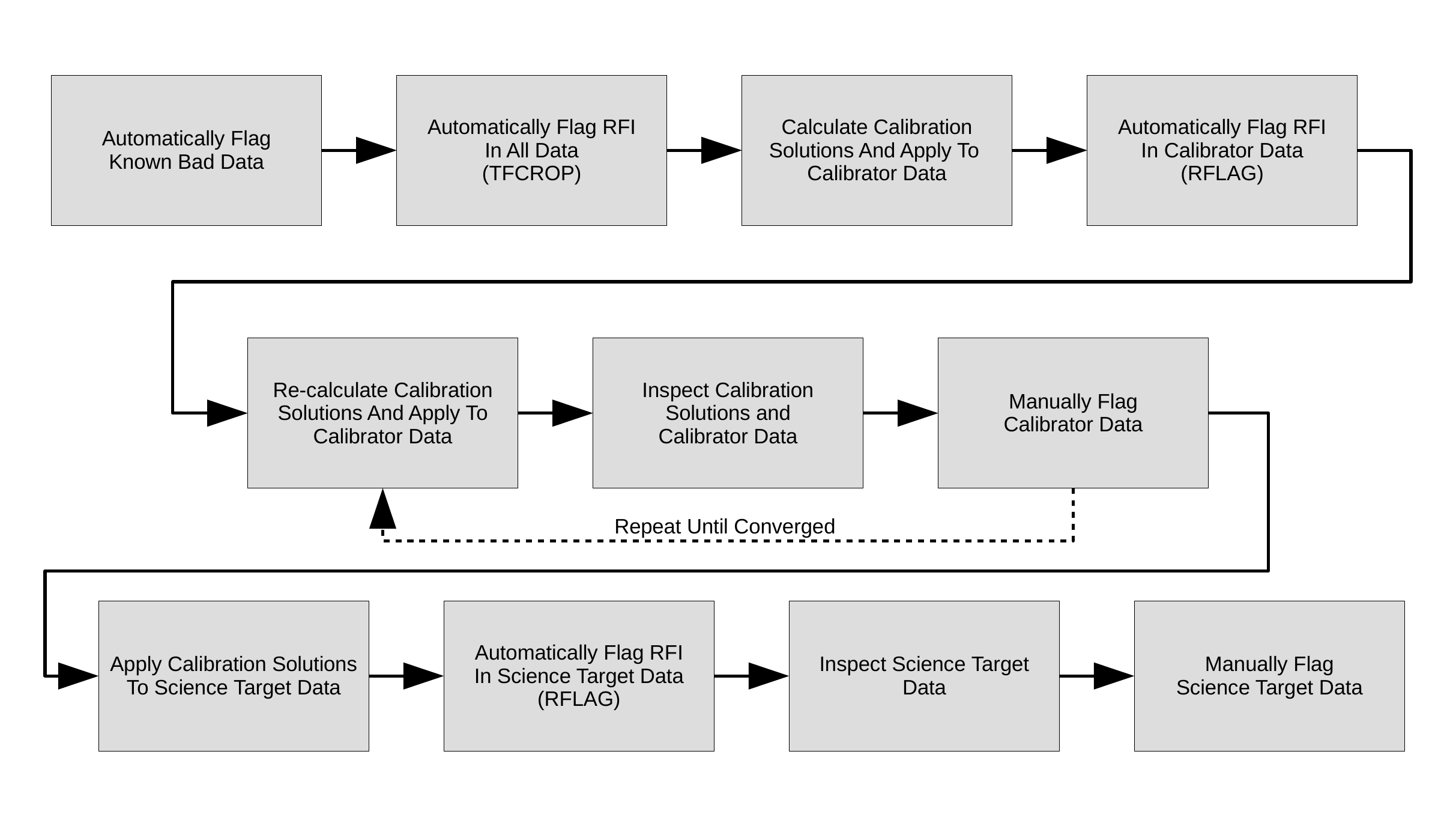}
  \caption{The steps performed by the WISP calibration pipeline.}
  \label{fig:calibration_flowchart}
\end{figure}

\subsection{Calibration}

The calibration pipeline in WISP handles all of the necessary
calibration steps: 1) flag bad data, 2) compute calibration solutions,
and 3) apply calibration solutions. Using a combination of automated
flagging and manual flagging, this pipeline is extremely
time-efficient. A summary of the calibration process is shown in
Figure~\ref{fig:calibration_flowchart}.

The raw data are contaminated by several sources of bad data,
including bad channels in the correlator (``birdies''), shadowed
antennas, the beginning and end of scans when the antennas are not on
source, and radio frequency interference (RFI). The pipeline begins by
flagging the known sources of bad data: birdies, shadowed antennas,
and off-source antennas.  We then use the automatic flagging algorithm
\textit{TFCROP}, as implemented in CASA, to catch any RFI. This task
is the recommended auto-flagging task to use on \textit{un-calibrated}
data by detecting outliers in the time-frequency domain. Our tests
show that \textit{TFCROP} is excellent at finding \({\sim}95\%\) of
bright, short-duration, limited-frequency RFI but often misses complex
RFI, which can completely compromise the data quality of a spectral
line window.

The data are now pruned of obvious RFI, but some low-level RFI may
still remain. Using the calibrator data, we compute preliminary
calibration solutions for the absolute flux, bandpass, delays, and
complex gains. We apply these calibration solutions to the
calibrators, thereby removing any instrumental effects from the
data. These calibrated data should be well-behaved and thus it is
easier to identify any RFI or otherwise bad data. Using the CASA
algorithm \textit{RFLAG}, we again automatically flag the calibrator
data to identify any remaining bad data. This algorithm differs from
\textit{TFCROP} in that it requires the data to be calibrated. It then
computes statistics on small chunks of the data and flags data using a
threshold determined by these statistics.  The \textit{RFLAG}
algorithm successfully flags most of the RFI missed by
\textit{TFCROP}, but it often misses other bad data, such as
misbehaving antennas or RFI compromising an entire spectral line
window. This algorithm will also flag the peak of our spectral lines
if the spectral line is very bright (\({\gsim}500\) mJy
beam\(^{-1}\)). In these cases, we must manually \textit{un-flag} the
spectral lines.

At this stage we have run the data through two independent automatic
flagging algorithms. We re-compute and re-apply the calibration
solutions. We then generate diagnostic plots to inspect the quality of
the calibrator data and calibration solutions. These diagnostic plots
are slices of the data in many dimensions: for example, complex
amplitude as a function of complex phase, real amplitude as a function
of time, and complex phase as a function of frequency. If we notice
any heretofore unidentified bad data, we manually flag it,
re-calculate and re-apply the calibration solutions, re-generate the
diagnostic plots, and re-inspect them. We iterate this process until
the calibrator data are free of any bad data.

After accurate calibration solutions are established, the science
target data can be calibrated. Using the final calibration solutions,
we calibrate the absolute flux, bandpass, delay, and complex gain of
the science target data using the secondary calibrator located closest
to the science target on the sky. We then run the \textit{RFLAG}
algorithm on the science target data to identify any RFI. Finally, we
generate diagnostic plots for the science target data and manually
flag any remaining bad data.

\begin{figure}
  \centering
  \includegraphics[width=\linewidth]{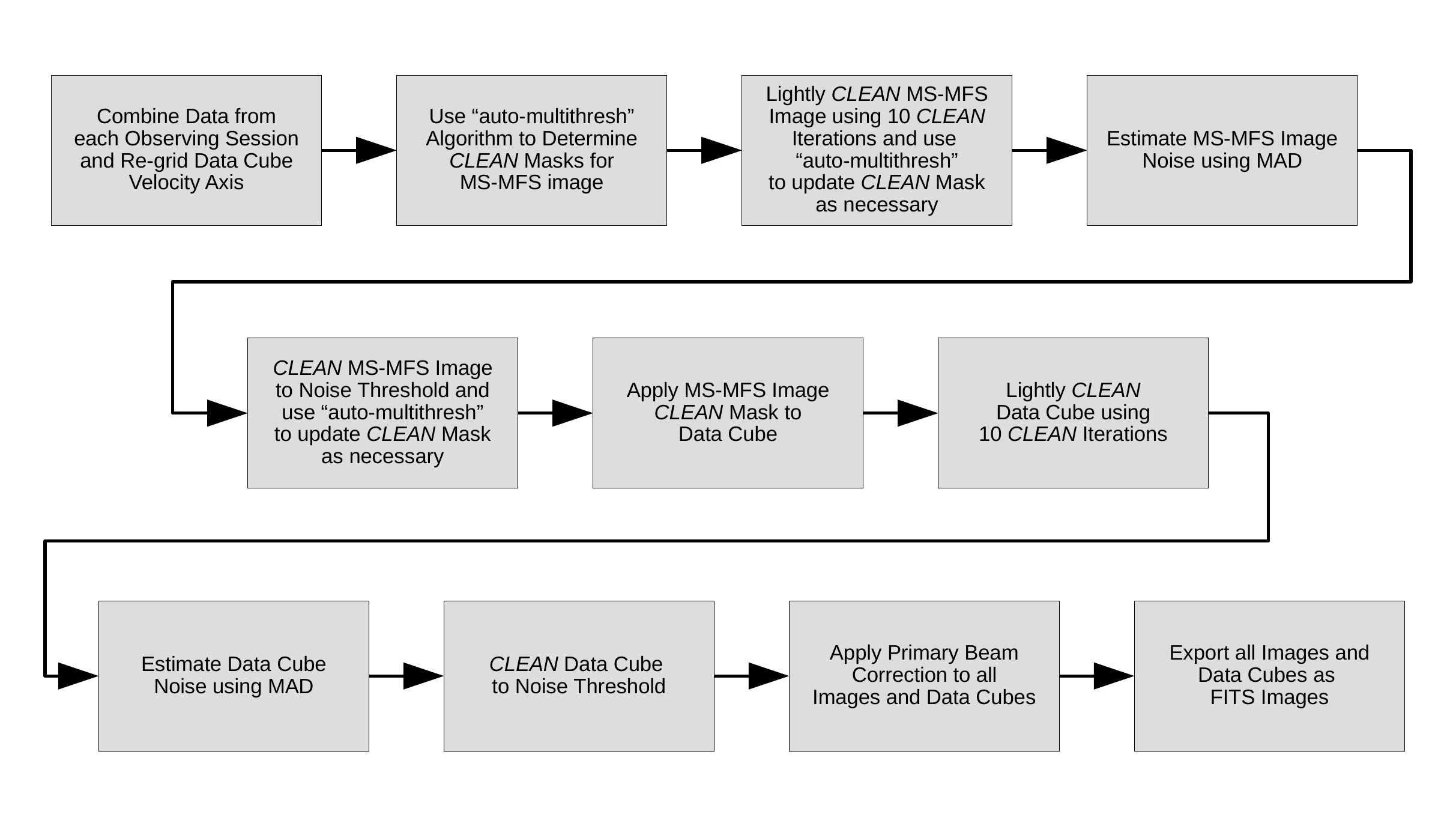}
  \caption{The steps performed by the WISP imaging pipeline for a
    single spectral line window.}
  \label{fig:imaging_flowchart}
\end{figure}

The final data products generated by the WISP calibration pipeline
are: 1) fully-calibrated visibilities, 2) a flag table identifying all
of the flagged data, 3) the final calibration solution tables, 4) the
final diagnostic plots, and 5) a copy of the fully-calibrated
visibilities for each individual field. We use WISP to calibrate each
observing session of the SHRDS. If a given field is observed in
multiple sessions, we combine all of the data for that field before
imaging.

\subsection{Imaging}

The imaging pipeline in WISP automatically creates several images for
each of our fields. This automation is necessary due to the large
amount of computing time required to image and \textit{CLEAN} each
field.  The \textit{CLEAN} and other imaging parameters are initially
set by the user and then applied to every image. All of the imaging
and \textit{CLEAN}ing is performed using the CASA task
\textit{TCLEAN}. We generate the following images for each field: 1) a
multi-scale, multi-frequency synthesis (MS-MFS) image produced by
combining the two 2\ghz\ continuum windows, 2) a MS-MFS image of each
2\ghz\ bandwidth continuum window, 3) MS-MFS images of each
64\mhz\ bandwidth spectral line window, and 4) multi-scale data cubes
of each spectral line window. Figure~\ref{fig:imaging_flowchart}
summarizes the imaging process for a single spectral line window.

\begin{figure}
  \centering
  \includegraphics[width=0.85\linewidth]{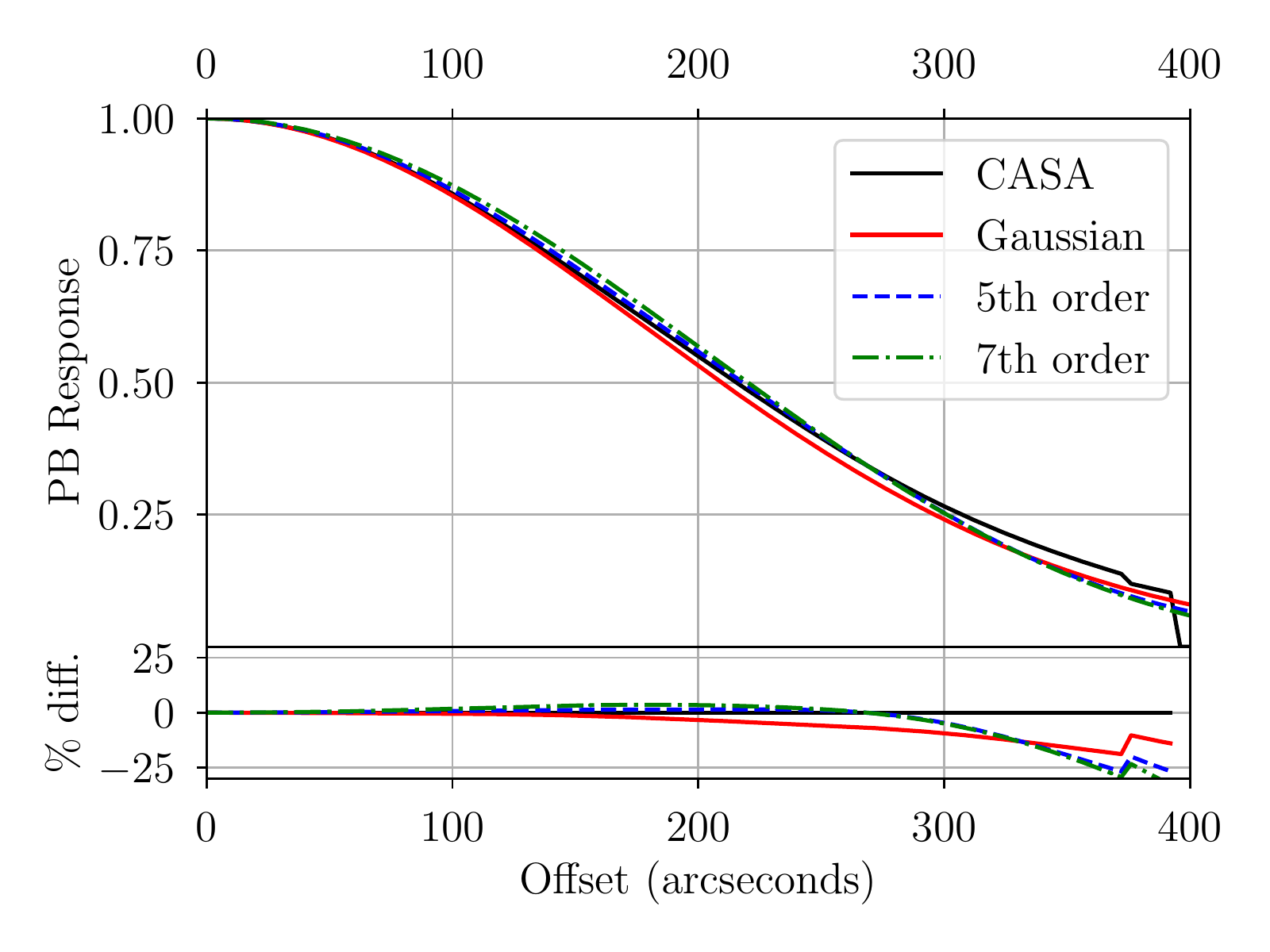}
  \caption{The primary beam response generated by \textit{TCLEAN}
    (black solid line) and three empirical models by
    \citet{wieringa1992}: a Gaussian (red solid line), a 5th order
    polynomial (blue dashed line), and a 7th order polynomial (green
    dash-dotted line). The bottom panel shows the percent difference
    between the three \citet{wieringa1992} models and the
    \textit{TCLEAN}-generated profile.}
  \label{fig:primary_beam}
\end{figure}

We use the newly-implemented ``auto-multithresh'' algorithm to
automatically determine \textit{CLEAN} masks in our images. This
algorithm applies user-defined thresholds to determine the locations
of ``real'' emission around which to place a \textit{CLEAN} mask. It
first masks any emission brighter than a given threshold above the
brightest sidelobe in the image, based on the
\textit{TCLEAN}-generated dirty beam image. After several iterations
of \textit{CLEAN}, the dirty beam sidelobes are suppressed and the
algorithm begins masking any emission brighter than a given threshold
above the estimated noise in the \textit{TCLEAN}-generated residual
image. The sidelobe masking threshold is very sensitive to the
\textit{uv-}coverage of the data; images with very complete
\textit{uv-}coverage, either because of large bandwidths or more hour
angle coverage, have dirty beams with lower sidelobes, whereas images
with poor \textit{uv-}coverage may have sidelobes comparable in gain
to the synthesized beam. Therefore, we use a separate set of
thresholds for the large-bandwidth continuum windows and the
small-bandwidth spectral line windows.

The ``auto-multithresh'' algorithm is time consuming when applied to
each channel of a data cube independently. We reduce the
\textit{CLEAN} computation time by applying the ``auto-multithresh''
\textit{CLEAN} mask from the MS-MFS image of a given spectral window
to all of the channels in the data cube for that spectral window.
This is possible because the morphology of our target \hii\ regions
and \hii\ region candidates should not change from channel to channel;
the free-free continuum and RRL emission originate in the same volume
of gas.

Each image and data cube is ``lightly-\textit{CLEAN}ed'' by running
only 10 iterations of \textit{CLEAN}. From these data we estimate the
image noise and \textit{CLEAN} stopping criterion (i.e. \textit{CLEAN}
threshold). We compute the median absolute deviation (MAD) of the
\textit{TCLEAN}-generated residual image to estimate the
\textit{CLEAN} threshold. The MAD is a robust estimator of the noise
of a data set even in the presence of outliers.  For Gaussian noise,
MAD is related to the root-mean-square, rms, by \({\rm rms} \simeq
1.4826\,{\rm MAD}\).  We use this rms as the stopping criterion for
\textit{CLEAN}: when the maximum absolute value of the residual image
within the \textit{CLEAN} mask is below this value, we stop
\textit{CLEAN}. This entire process usually takes fewer than 50
iterations of \textit{CLEAN} for the MS-MFS images and 200 iterations
for the spectral data cubes.

In order to average together each spectral window into a final
\(<\)H\(n\alpha\)\(>\) spectrum, we must re-grid the spectral windows
to a common velocity frame. Using the CASA task \textit{CVEL2}, we
linearly interpolate each spectral line window to the kinematic local
standard of rest (LSRK) reference frame in 321 velocity channels with
a \(2.5\kms\) width starting at \(-400\kms\).  This gives us velocity
coverage from \(\vlsr = -400\kms\) to \(+400\kms\), sufficient to
detect all Galactic \hii\ regions.

Finally, each image and data cube is primary beam corrected. We divide
each image and cube by the \textit{TCLEAN}-generated primary beam
image. For our full 4\ghz\ bandwidth continuum images, we use the CASA
task \textit{WIDEBANDPBCOR} which considers the varying primary beam
over a large fractional bandwidth (\(4-10\ghz\) in our case).  For all
other images, we use \textit{IMPBCOR} which applies the primary beam
correction using the primary beam of the center frequency.
Figure~\ref{fig:primary_beam} shows the \textit{TCLEAN}-generated
primary beam profile for a 4\ghz\ bandwidth MS-MFS image. We also show
the Gaussian, 5th order polynomial, and 7th order polynomial empirical
primary beam models from \citet{wieringa1992}. The CASA primary beam
deviates significantly (\(\gsim10\%\)) from these models beyond
\({\sim}300\,\text{arcseconds}\) from the primary beam center.  Thus
the continuum and RRL intensities of sources at the edges of an
observed field are inaccurate.

The final data products generated by the WISP imaging pipeline for
each field are: 1) a MS-MFS image of the combined 4\ghz\ bandwidth
continuum windows, 2) a MS-MFS image of each 2\ghz\ bandwidth
continuum window, 3) MS-MFS images of each 64\mhz\ bandwidth spectral
line window, and 4) multi-scale data cubes of each spectral line
window. These images are saved both as CASA and FITS images. All
subsequent analysis is performed using the FITS images, so the CASA
images are deleted.
  
\section{Multiple Detections}\label{sec:app_multiple_detections}

A \textit{WISE} Catalog source may appear in multiple fields.  The
data with the highest quality factor and/or largest signal-to-noise
ratio is included in the main catalogs
(Tables~\ref{tab:cont_notap},~\ref{tab:cont_uvtap},~\ref{tab:rrl_notap},~and~\ref{tab:rrl_uvtap}).
Here we give the radio continuum and \(<\)H\(n\alpha\)\(>\) RRL
properties for these ``multiple detections'' as measured in each
independent field. Most of these sources are located near the edge of
the field, so the continuum and RRL intensities may be inaccurate (see
Appendix~\ref{sec:app_data_reduction}). The data in the main catalogs,
not the multiple detections catalogs, should be used for all
subsequent analysis. In the next data release we will combine the data
from each field to improve our sensitivity and to measure the
continuum and RRL properties of these sources more accurately.

The continuum properties of the multiple detections are listed in
Table~\ref{tab:cont_notap_mult} (non-tapered) and
Table~\ref{tab:cont_uvtap_mult} (\textit{uv}-tapered). For each
source, we list the \textit{WISE} Catalog name, position of the peak
radio continuum emission, fields containing the source, epoch,
synthesized beam area in the continuum image, separation between the
observed continuum peak position and infrared position,
\(\Delta\theta\), MFS-synthesized frequency of the continuum image,
\(\nu_C\), peak continuum flux density, \(S_C\), rms noise, rms\(_C\),
and a quality factor, QF. These tables contain one row for each field
in which the source is found.

The \(<\)H\(n\alpha\)\(>\) RRL properties of the multiple detections
are listed in Table~\ref{tab:rrl_notap_mult} (non-tapered) and
Table~\ref{tab:rrl_uvtap_mult} (\textit{uv}-tapered). For each source,
we give the \textit{WISE} Catalog name, field, epoch, weighted-average
frequency of the \(<\)H\(n\alpha\)\(>\) RRL, \(\nu_L\), Gaussian fits
to the peak intensity, \(S_L\), LSR velocity, \(V_{\rm LSR}\), and
FWHM line width, \(\Delta V\), rms spectral noise in the line-free
region of the spectrum, rms\(_L\), signal-to-noise ratio, S/N, and a
quality factor, QF. These tables also contain one row for each field
in which the source is found.

\newpage

\begin{longrotatetable}
\begin{deluxetable*}{lcccccccrrrc}
\centering
\tablewidth{0pt}
\tabletypesize{\scriptsize}
\tablecaption{Bright Catalog Fields and Targets\label{tab:target_sample}}
\tablehead{
\colhead{Field} & \colhead{RA}         & \colhead{Dec.}       & \colhead{Epoch} & \colhead{Target\tablenotemark{a}} & \colhead{Catalog\tablenotemark{b}} & \colhead{RA}         & \colhead{Dec.}       & \colhead{\(R_{\rm IR}\)}   & \colhead{\(S_{6\ghz}\)}       & \colhead{\(\Delta\theta\)\tablenotemark{c}} & \colhead{Author} \\
\colhead{}      & \colhead{J2000}      & \colhead{J2000}      & \colhead{}      & \colhead{}                        & \colhead{}                         & \colhead{J2000}      & \colhead{J2000}      & \colhead{(arcsec)}         & \colhead{(mJy}                & \colhead{(arcsec)}                          & \colhead{}      \\
\colhead{}      & \colhead{(hh:mm:ss)} & \colhead{(dd:mm:ss)} & \colhead{}      & \colhead{}                        & \colhead{}                         & \colhead{(hh:mm:ss)} & \colhead{(dd:mm:ss)} & \colhead{}                 & \colhead{beam\(^{-1}\))}      & \colhead{}                                  & \colhead{}      
}
\startdata
shrds030 & 08:26:18.8 & $-40$:48:36.5 & 2016 & G259.013$-$01.546 & Q & 08:26:13.5 & $-40$:46:45.0 & $68.08$ & \nodata & \nodata &  \\
 & & & & G259.057$-$01.544$^T$ & C & 08:26:22.1 & $-40$:48:49.4 & $103.98$ & $135.54$ & $39.81$ &  \\
 & & & & G259.086$-$01.612 & C & 08:26:09.9 & $-40$:52:36.1 & $103.98$ & \nodata & \nodata &  \\
shrds035 & 08:37:05.6 & $-39$:44:51.8 & 2015 & G259.359+00.820 & C & 08:37:17.6 & $-39$:39:30.3 & $500.76$ & $94.64$ & $350.65$ &  \\
 & & & & G259.771+00.541 & C & 08:37:25.8 & $-40$:09:15.3 & $3361.35$ & \nodata & \nodata &  \\
shrds060 & 08:48:40.9 & $-42$:53:10.3 & 2015 & G263.237+00.509$^T$ & C & 08:48:44.0 & $-42$:54:01.6 & $204.56$ & $168.93$ & $61.36$ &  \\
shrds063 & 08:49:49.2 & $-43$:09:37.3 & 2015 & G263.554+00.656 & C & 08:50:27.3 & $-43$:03:08.8 & $1043.51$ & $43.67$ & $568.91$ &  \\
shrds073 & 08:52:41.6 & $-44$:13:27.4 & 2015 & G264.220+00.216 & C & 08:50:54.8 & $-43$:50:47.4 & $1824.13$ & $42.02$ & $526.47$ &  \\
\enddata
\tablenotetext{a}{``T'' indicates that this source is a ``nominal'' target (see text)}
\tablenotetext{b}{The \textit{WISE} Catalog designation: ``K'' is known \hii\ region, ``C'' is \hii\ region candidate, ``Q'' is radio-quiet \hii\ region candidate, and ``G'' is \hii\ candidate associated with an \hii\ region group.}
\tablenotetext{c}{The separation between the \textit{WISE} Catalog infrared position and the position of the SUMSS continuum peak.}
\tablerefs{(GBT HRDS) \citet{anderson2011,anderson2015b,anderson2018}; (CH87) \citet{caswell1987}; (L89) \citet{lockman1989}; (W70) \citet{wilson1970}}
\end{deluxetable*}
\end{longrotatetable}

\begin{deluxetable*}{lll}
\centering
\tablewidth{0pt}
\tabletypesize{\normalsize}
\tablecaption{Observation Summary\label{tab:observations}}
\tablehead{
  \colhead{Parameters} & \colhead{Epoch: 2015} & \colhead{Epoch: 2016}
}
\startdata
H75 Observing Dates & 2015$-$07$-$24 to 2015$-$08$-$23 & 2016$-$07$-$23 to 2016$-$08$-$15 \\
H75 Observing Time (hr) & 121 & 150 \\
\hline
H168 Observing Dates & 2015$-$09$-$18 to 2015$-$10$-$14 & 2016$-$09$-$03 to 2016$-$09$-$29 \\
H168 Observing Time (hr) & 63 & 144 \\
\hline
Total Observing Time (hr) & 184 & 294 \\
\hline
Primary Calibrators & 0823$-$500, 1934$-$638 & 0823$-$500, 1934$-$638 \\
Secondary Calibrators & 0906$-$47, 1036$-$52, & 1036$-$52, 1613$-$586, \\
                      & 1613$-$586, 1714$-$336, & j1322$-$6532 \\
                      & 1714$-$397, j1322$-$6532 & \\
\enddata
\end{deluxetable*}

\begin{deluxetable*}{lcccccc}
\centering
\tablewidth{0pt}
\tabletypesize{\footnotesize}
\tablecaption{Correlator Setup\label{tab:backend}}
\tablehead{
  \colhead{Window} & \colhead{$\nu_{\rm center}$} & \colhead{Bandwidth} & \colhead{Channels} & \colhead{$\Delta\nu$} & \colhead{RRL} & \colhead{$\nu_{\rm RRL}$} \\
  \colhead{}       & \colhead{(MHz)}  & \colhead{(MHz)}     & \colhead{}         & \colhead{(kHz)}   & \colhead{}    & \colhead{(MHz)}
}
\startdata
0 & 5505 & 2112 & 33 & 64000 & \dots & \dots \\
1 & 4609 & 64 & 2049 & 31.25 & H112\(\alpha\) & 4618.790 \\
2 & 4737 & 64 & 2049 & 32.25 & H111\(\alpha\) & 4744.184 \\
3 & 4865 & 64 & 2049 & 31.25 & H110\(\alpha\) & 4874.158 \\
4 & 4993 & 64 & 2049 & 31.25 & H109\(\alpha\) & 5008.924 \\
5 & 5153 & 64 & 2049 & 31.25 & H108\(\alpha\) & 5148.704 \\
6 & 5281 & 64 & 2049 & 31.25 & H107\(\alpha\) & 5293.733 \\
7 & 5441 & 64 & 2049 & 31.25 & H106\(\alpha\) & 5444.262 \\
8 & 5601 & 64 & 2049 & 31.25 & H105\(\alpha\) & 5600.551 \\
9 & 5761 & 64 & 2049 & 31.25 & H104\(\alpha\) & 5762.881 \\
10 & 5921 & 64 & 2049 & 31.25 & H103\(\alpha\) & 5931.546 \\
11 & 6113 & 64 & 2049 & 31.25 & H102\(\alpha\) & 6106.857 \\
12 & 6305 & 64 & 2049 & 31.25 & H101\(\alpha\) & 6289.145 \\
13 & 6465 & 64 & 2049 & 31.25 & H100\(\alpha\) & 6478.761 \\
14 & 8540 & 2112 & 33 & 64000 & \dots & \dots \\
15 & 7548 & 64 & 2049 & 31.25 & H95\(\alpha\) & 7550.616 \\
16 & 7804 & 64 & 2049 & 31.25 & H94\(\alpha\) & 7792.872 \\
17 & 8060 & 64 & 2049 & 31.25 & H93\(\alpha\) & 8045.604 \\
18 & 8316 & 64 & 2049 & 31.25 & H92\(\alpha\) & 8309.384 \\
19 & 8572 & 64 & 2049 & 31.25 & H91\(\alpha\) & 8584.823 \\
20 & 9180 & 64 & 2049 & 31.25 & H89\(\alpha\) & 9173.323 \\
21 & 8500 & 64 & 2049 & 31.25 & H88\(\alpha\) & 9487.823 \\
\enddata
\end{deluxetable*}

\newpage

\begin{longrotatetable}
\begin{deluxetable*}{lccccrrcrrc}
\centering
\tablewidth{0pt}
\tabletypesize{\scriptsize}
\tablecaption{Non-Tapered Image Continuum Properties\label{tab:cont_notap}}
\tablehead{
\colhead{Target} & \colhead{RA}         & \colhead{Dec.}       & \colhead{Field} & \colhead{Epoch\tablenotemark{a}}  & \colhead{Beam}           & \colhead{\(\Delta\theta\)\tablenotemark{b}} & \colhead{\(\nu_C\)} & \colhead{$S_C$}           & \colhead{rms$_C$}       & \colhead{QF} \\
\colhead{}       & \colhead{J2000}      & \colhead{J2000}      & \colhead{}      & \colhead{}                        & \colhead{Area}           & \colhead{}                                  & \colhead{}          & \colhead{}                & \colhead{}              & \colhead{} \\
\colhead{}       & \colhead{(hh:mm:ss)} & \colhead{(dd:mm:ss)} & \colhead{}      & \colhead{}                        & \colhead{(arcsec\(^2\))} & \colhead{(arcsec)}                          & \colhead{(MHz)}     & \colhead{(mJy}            & \colhead{(mJy}          & \colhead{} \\
\colhead{}       & \colhead{}           & \colhead{}           & \colhead{}      & \colhead{}                        & \colhead{}               & \colhead{}                                  & \colhead{}          & \colhead{beam\(^{-1}\))} & \colhead{beam\(^{-1}\))} & \colhead{} 
}
\startdata
G213.833+00.618 & 06:55:16.9 & $+0$:31:12.3 & g213.833+00.618 & 2014* & \nodata & 2.5 & \nodata & \nodata & \nodata & \nodata \\ 
G230.354$-$00.597 & 07:21:49.5 & $-15$:44:09.9 & g230.354-00.597 & 2014* & \nodata & 2.2 & \nodata & \nodata & \nodata & \nodata \\ 
G259.057$-$01.544 & 08:26:18.1 & $-40$:48:36.0 & shrds030 & 2016 & 1353 & 47.4 & 7022 & 69.76 & 0.31 & B \\ 
G259.359+00.820 & 08:37:05.3 & $-39$:44:47.8 & shrds035 & 2015 & 2501 & 347.8 & 7022 & 19.62 & 0.06 & A \\ 
G263.237+00.509 & 08:48:37.6 & $-42$:53:54.1 & shrds060 & 2015 & 3077 & 70.1 & 7022 & 67.53 & 0.43 & B \\ 
G264.681+00.272 & 08:52:55.7 & $-44$:16:11.3 & shrds073 & 2015 & 1382 & 385.2 & 7022 & 2.94 & 0.23 & C \\ 
G265.191$-$00.641 & 08:50:38.3 & $-45$:08:20.4 & shrds075 & 2015 & 2479 & 17.1 & 7022 & 56.13 & 0.31 & B \\ 
G267.730$-$01.100 & 08:58:04.5 & $-47$:22:51.2 & shrds078 & 2016 & 1627 & 12.2 & 7023 & 226.76 & 8.29 & A \\
\enddata
\tablenotetext{a}{Rows with epochs including an asterisk (*) are copied directly from \citet{brown2017}. They did not give the continuum fluxes for their detections.}
\tablenotetext{b}{The separation between the \textit{WISE} Catalog infrared position and the position of the SHRDS continuum peak.}
\end{deluxetable*}
\end{longrotatetable}

\newpage

\begin{longrotatetable}
\begin{deluxetable*}{lccccrrcrrc}
\centering
\tablewidth{0pt}
\tabletypesize{\scriptsize}
\tablecaption{\textit{uv-}Tapered Image Continuum Properties\label{tab:cont_uvtap}}
\tablehead{
\colhead{Target} & \colhead{RA}         & \colhead{Dec.}       & \colhead{Field} & \colhead{Epoch\tablenotemark{a}}  & \colhead{Beam}           & \colhead{\(\Delta\theta\)\tablenotemark{b}} & \colhead{\(\nu_C\)} & \colhead{$S_C$}           & \colhead{rms$_C$}       & \colhead{QF} \\
\colhead{}       & \colhead{J2000}      & \colhead{J2000}      & \colhead{}      & \colhead{}                        & \colhead{Area}           & \colhead{}                                  & \colhead{}          & \colhead{}                & \colhead{}              & \colhead{} \\
\colhead{}       & \colhead{(hh:mm:ss)} & \colhead{(dd:mm:ss)} & \colhead{}      & \colhead{}                        & \colhead{(arcsec\(^2\))} & \colhead{(arcsec)}                          & \colhead{(MHz)}     & \colhead{(mJy}            & \colhead{(mJy}          & \colhead{} \\
\colhead{}       & \colhead{}           & \colhead{}           & \colhead{}      & \colhead{}                        & \colhead{}               & \colhead{}                                  & \colhead{}          & \colhead{beam\(^{-1}\))} & \colhead{beam\(^{-1}\))} & \colhead{} 
}
\startdata
G213.833+00.618 & 06:55:16.9 & $+0$:31:12.3 & g213.833+00.618 & 2014* & \nodata & 2.5 & \nodata & \nodata & \nodata & \nodata \\ 
G230.354$-$00.597 & 07:21:49.5 & $-15$:44:09.9 & g230.354-00.597 & 2014* & \nodata & 2.2 & \nodata & \nodata & \nodata & \nodata \\ 
G259.057$-$01.544 & 08:26:19.2 & $-40$:48:32.2 & shrds030 & 2016 & 9022 & 37.8 & 7022 & 192.46 & 0.98 & B \\ 
G259.359+00.820 & 08:37:05.6 & $-39$:44:47.7 & shrds035 & 2015 & 9230 & 346.0 & 7022 & 20.98 & 0.13 & A \\ 
G263.237+00.509 & 08:48:38.4 & $-42$:53:58.1 & shrds060 & 2015 & 9205 & 61.8 & 7022 & 149.64 & 0.89 & A \\ 
G264.681+00.272 & 08:52:50.5 & $-44$:13:51.4 & shrds073 & 2015 & 8244 & 237.1 & 7022 & 4.91 & 1.88 & C \\ 
G265.191$-$00.641 & 08:50:38.6 & $-45$:08:12.4 & shrds075 & 2015 & 9233 & 19.1 & 7022 & 61.11 & 0.73 & B \\ 
G267.730$-$01.100 & 08:58:04.4 & $-47$:22:55.2 & shrds078 & 2016 & 8498 & 15.1 & 7023 & 404.28 & 32.87 & A \\
\enddata
\tablenotetext{a}{Rows with epochs including an asterisk (*) are copied directly from \citet{brown2017}. They did not give the continuum fluxes for their detections.}
\tablenotetext{b}{The separation between the \textit{WISE} Catalog infrared position and the position of the SHRDS continuum peak.}
\end{deluxetable*}
\end{longrotatetable}

\newpage

\begin{longrotatetable}
\begin{deluxetable*}{lcccrrrrrc}
\centering
\tablewidth{0pt}
\tabletypesize{\scriptsize}
\tablecaption{Non-Tapered Image RRL Properties\label{tab:rrl_notap}}
\tablehead{
\colhead{Target} & \colhead{Field} & \colhead{Epoch\tablenotemark{a}}   & \colhead{\(\nu_L\)\tablenotemark{b}} & \colhead{$S_L$}          & \colhead{$V_{\rm LSR}$} & \colhead{$\Delta V$}    & \colhead{rms$_L$}      & \colhead{S/N} & \colhead{QF} \\
\colhead{}       & \colhead{}      & \colhead{}                         & \colhead{(MHz)}                      & \colhead{(mJy}           & \colhead{(km s$^{-1}$)} & \colhead{(km s$^{-1}$)} & \colhead{(mJy}         & \colhead{}    & \colhead{} \\
\colhead{}       & \colhead{}      & \colhead{}                         & \colhead{}                           & \colhead{beam\(^{-1}\))} & \colhead{}              & \colhead{}              & \colhead{beam$^{-1}$)} & \colhead{}    & \colhead{} 
}
\startdata
G213.833+00.618 & g213.833+00.618 & 2014* & \nodata & 1.90 $\pm$ 2.30 & 53.20 $\pm$ 20.20 & 33.50 $\pm$ 20.90 & 1.40 & 3.5 & C \\ 
G230.354$-$00.597 & g230.354-00.597 & 2014* & \nodata & 2.10 $\pm$ 2.00 & 69.00 $\pm$ 16.10 & 35.00 $\pm$ 16.70 & 1.50 & 3.7 & C \\ 
G259.057$-$01.544 & shrds030 & 2016 & 6474 & 5.20 $\pm$ 0.40 & 59.20 $\pm$ 1.10 & 26.80 $\pm$ 2.60 & 1.10 & 10.9 & B \\ 
G263.237+00.509 & shrds060 & 2015 & 6338 & 8.10 $\pm$ 1.10 & 10.40 $\pm$ 1.20 & 17.50 $\pm$ 2.70 & 1.90 & 7.8 & C \\ 
G267.730$-$01.100 & shrds078 & 2016 & 6928 & 26.50 $\pm$ 0.70 & 7.30 $\pm$ 0.30 & 21.50 $\pm$ 0.60 & 1.50 & 36.6 & A \\ 
G269.068$-$01.114 & shrds085 & 2016 & 6028 & 75.70 $\pm$ 1.50 & 13.20 $\pm$ 0.20 & 25.00 $\pm$ 0.60 & 3.40 & 49.4 & A \\ 
G269.159$-$01.119 & shrds085 & 2016 & 6513 & 28.80 $\pm$ 0.70 & 10.70 $\pm$ 0.30 & 25.90 $\pm$ 0.80 & 1.80 & 36.3 & A \\ 
G269.167$-$01.357 & shrds086 & 2016 & 6445 & 7.30 $\pm$ 0.50 & 2.80 $\pm$ 0.80 & 23.90 $\pm$ 1.90 & 1.00 & 16.1 & A \\
\enddata
\tablenotetext{a}{Rows with epochs including an asterisk (*) are copied directly from \citet{brown2017}. They did not give the weigted average frequency of their Hn\(\alpha\) RRL spectra.}
\tablenotetext{b}{This is the weighted average frequency of the \(<\)H\(n\alpha\)\(>\) stacked spectrum.}
\end{deluxetable*}
\end{longrotatetable}

\newpage

\begin{longrotatetable}
\begin{deluxetable*}{lcccrrrrrc}
\centering
\tablewidth{0pt}
\tabletypesize{\scriptsize}
\tablecaption{\textit{uv}-Tapered Image RRL Properties\label{tab:rrl_uvtap}}
\tablehead{
\colhead{Target} & \colhead{Field} & \colhead{Epoch\tablenotemark{a}}   & \colhead{\(\nu_L\)\tablenotemark{b}} & \colhead{$S_L$}          & \colhead{$V_{\rm LSR}$} & \colhead{$\Delta V$}    & \colhead{rms$_L$}      & \colhead{S/N} & \colhead{QF} \\
\colhead{}       & \colhead{}      & \colhead{}                         & \colhead{(MHz)}                      & \colhead{(mJy}           & \colhead{(km s$^{-1}$)} & \colhead{(km s$^{-1}$)} & \colhead{(mJy}         & \colhead{}    & \colhead{} \\
\colhead{}       & \colhead{}      & \colhead{}                         & \colhead{}                           & \colhead{beam\(^{-1}\))} & \colhead{}              & \colhead{}              & \colhead{beam$^{-1}$)} & \colhead{}    & \colhead{} 
}
\startdata
G213.833+00.618 & g213.833+00.618 & 2014* & \nodata & 1.90 $\pm$ 2.30 & 53.20 $\pm$ 20.20 & 33.50 $\pm$ 20.90 & 1.40 & 3.5 & C \\ 
G230.354$-$00.597 & g230.354-00.597 & 2014* & \nodata & 2.10 $\pm$ 2.00 & 69.00 $\pm$ 16.10 & 35.00 $\pm$ 16.70 & 1.50 & 3.7 & C \\ 
G259.057$-$01.544 & shrds030 & 2016 & 6295 & 10.80 $\pm$ 0.80 & 58.00 $\pm$ 0.90 & 24.50 $\pm$ 2.10 & 2.10 & 11.3 & B \\ 
G263.237+00.509 & shrds060 & 2015 & 6196 & 16.90 $\pm$ 1.30 & 11.10 $\pm$ 0.60 & 17.20 $\pm$ 1.50 & 2.30 & 13.5 & B \\ 
G267.730$-$01.100 & shrds078 & 2016 & 6447 & 32.40 $\pm$ 1.30 & 7.70 $\pm$ 0.40 & 22.70 $\pm$ 1.00 & 2.30 & 29.6 & A \\ 
G269.068$-$01.114 & shrds085 & 2016 & 6127 & 149.70 $\pm$ 3.20 & 13.60 $\pm$ 0.20 & 23.70 $\pm$ 0.60 & 6.10 & 52.8 & A \\ 
G269.159$-$01.119 & shrds085 & 2016 & 6411 & 53.40 $\pm$ 1.50 & 11.00 $\pm$ 0.30 & 23.60 $\pm$ 0.80 & 3.10 & 37.1 & A \\ 
G269.167$-$01.357 & shrds086 & 2016 & 6515 & 22.00 $\pm$ 1.00 & 3.20 $\pm$ 0.50 & 22.10 $\pm$ 1.20 & 2.10 & 22.3 & A \\
\enddata
\tablenotetext{a}{Rows with epochs including an asterisk (*) are copied directly from \citet{brown2017}. They did not give the weigted average frequency of their Hn\(\alpha\) RRL spectra.}
\tablenotetext{b}{This is the weighted average frequency of the \(<\)H\(n\alpha\)\(>\) stacked spectrum.}
\end{deluxetable*}
\end{longrotatetable}

\newpage

\begin{longrotatetable}
\begin{deluxetable*}{lccccrrcrrc}
\centering
\tablewidth{0pt}
\tabletypesize{\scriptsize}
\tablecaption{Non-Tapered Image Continuum Properties of Multiple Detections\label{tab:cont_notap_mult}}
\tablehead{
\colhead{Target} & \colhead{RA}         & \colhead{Dec.}       & \colhead{Field} & \colhead{Epoch\tablenotemark{a}}  & \colhead{Beam}           & \colhead{\(\Delta\theta\)\tablenotemark{b}} & \colhead{\(\nu_C\)} & \colhead{$S_C$}           & \colhead{rms$_C$}       & \colhead{QF} \\
\colhead{}       & \colhead{J2000}      & \colhead{J2000}      & \colhead{}      & \colhead{}                        & \colhead{Area}           & \colhead{}                                  & \colhead{}          & \colhead{}                & \colhead{}              & \colhead{} \\
\colhead{}       & \colhead{(hh:mm:ss)} & \colhead{(dd:mm:ss)} & \colhead{}      & \colhead{}                        & \colhead{(arcsec\(^2\))} & \colhead{(arcsec)}                          & \colhead{(MHz)}     & \colhead{(mJy}            & \colhead{(mJy}          & \colhead{} \\
\colhead{}       & \colhead{}           & \colhead{}           & \colhead{}      & \colhead{}                        & \colhead{}               & \colhead{}                                  & \colhead{}          & \colhead{beam\(^{-1}\))}  & \colhead{beam\(^{-1}\))} & \colhead{} 
}
\startdata
G289.944$-$00.889 & 11:01:11.1 & $-60$:56:48.0 & shrds191 & 2015 & 2502 & 10.2 & 7023 & 123.68 & 1.47 & A \\ 
G289.944$-$00.889 & 11:01:09.7 & $-60$:56:40.8 & g290.012-00.867 & 2014 & 5394 & 16.4 & 6050 & 135.43 & 7.10 & C \\ 
G290.012$-$00.867 & 11:01:59.2 & $-60$:56:48.8 & g290.012-00.867 & 2014 & 5394 & 106.4 & 6050 & 200.17 & 3.66 & B \\ 
G290.012$-$00.867 & 11:01:45.4 & $-60$:57:25.8 & g290.012-00.867 & 2014* & \nodata & 0.8 & \nodata & \nodata & \nodata & \nodata \\ 
G290.323$-$02.984 & 10:56:33.0 & $-63$:00:43.9 & g290.323-02.984 & 2014 & 5453 & 26.8 & 6050 & 28.85 & 0.34 & A \\ 
G290.323$-$02.984 & 10:56:36.4 & $-63$:00:35.0 & g290.323-02.984 & 2014* & \nodata & 2.9 & \nodata & \nodata & \nodata & \nodata \\ 
G290.385$-$01.042 & 11:03:59.6 & $-61$:16:04.4 & g290.385-01.042 & 2014 & 5384 & 2.0 & 6050 & 23.30 & 0.87 & A \\ 
G290.385$-$01.042 & 11:03:59.4 & $-61$:16:06.7 & g290.385-01.042 & 2014* & \nodata & 0.6 & \nodata & \nodata & \nodata & \nodata \\
\enddata
\tablenotetext{a}{Rows with epochs including an asterisk (*) are copied directly from \citet{brown2017}. They did not give the continuum fluxes for their detections.}
\tablenotetext{b}{The separation between the \textit{WISE} Catalog infrared position and the position of the SHRDS continuum peak.}
\end{deluxetable*}
\end{longrotatetable}

\newpage

\begin{longrotatetable}
\begin{deluxetable*}{lccccrrcrrc}
\centering
\tablewidth{0pt}
\tabletypesize{\scriptsize}
\tablecaption{\textit{uv}-Tapered Image Continuum Properties of Multiple Detections\label{tab:cont_uvtap_mult}}
\tablehead{
\colhead{Target} & \colhead{RA}         & \colhead{Dec.}       & \colhead{Field} & \colhead{Epoch\tablenotemark{a}}  & \colhead{Beam}           & \colhead{\(\Delta\theta\)\tablenotemark{b}} & \colhead{\(\nu_C\)} & \colhead{$S_C$}           & \colhead{rms$_C$}       & \colhead{QF} \\
\colhead{}       & \colhead{J2000}      & \colhead{J2000}      & \colhead{}      & \colhead{}                        & \colhead{Area}           & \colhead{}                                  & \colhead{}          & \colhead{}                & \colhead{}              & \colhead{} \\
\colhead{}       & \colhead{(hh:mm:ss)} & \colhead{(dd:mm:ss)} & \colhead{}      & \colhead{}                        & \colhead{(arcsec\(^2\))} & \colhead{(arcsec)}                          & \colhead{(MHz)}     & \colhead{(mJy}            & \colhead{(mJy}          & \colhead{} \\
\colhead{}       & \colhead{}           & \colhead{}           & \colhead{}      & \colhead{}                        & \colhead{}               & \colhead{}                                  & \colhead{}          & \colhead{beam\(^{-1}\))}  & \colhead{beam\(^{-1}\))} & \colhead{} 
}
\startdata
G289.944$-$00.889 & 11:01:11.1 & $-60$:56:48.0 & shrds191 & 2015 & 9232 & 10.1 & 7023 & 145.89 & 3.69 & A \\ 
G289.944$-$00.889 & 11:01:09.7 & $-60$:56:44.3 & g290.012-00.867 & 2014 & 7104 & 13.1 & 6050 & 131.86 & 10.48 & B \\ 
G290.012$-$00.867 & 11:01:59.1 & $-60$:56:49.2 & g290.012-00.867 & 2014 & 7104 & 105.8 & 6050 & 196.21 & 5.43 & B \\ 
G290.012$-$00.867 & 11:01:45.4 & $-60$:57:25.8 & g290.012-00.867 & 2014* & \nodata & 0.8 & \nodata & \nodata & \nodata & \nodata \\ 
G290.323$-$02.984 & 10:56:33.0 & $-63$:00:44.2 & g290.323-02.984 & 2014 & 7202 & 27.2 & 6050 & 29.11 & 0.46 & A \\ 
G290.323$-$02.984 & 10:56:36.4 & $-63$:00:35.0 & g290.323-02.984 & 2014* & \nodata & 2.9 & \nodata & \nodata & \nodata & \nodata \\ 
G290.385$-$01.042 & 11:03:59.5 & $-61$:16:04.9 & g290.385-01.042 & 2014 & 7084 & 1.4 & 6050 & 24.97 & 1.35 & A \\ 
G290.385$-$01.042 & 11:03:59.4 & $-61$:16:06.7 & g290.385-01.042 & 2014* & \nodata & 0.6 & \nodata & \nodata & \nodata & \nodata \\
\enddata
\tablenotetext{a}{Rows with epochs including an asterisk (*) are copied directly from \citet{brown2017}. They did not give the continuum fluxes for their detections.}
\tablenotetext{b}{The separation between the \textit{WISE} Catalog infrared position and the position of the SHRDS continuum peak.}
\end{deluxetable*}
\end{longrotatetable}

\newpage

\begin{longrotatetable}
\begin{deluxetable*}{lcccrrrrrc}
\centering
\tablewidth{0pt}
\tabletypesize{\scriptsize}
\tablecaption{Non-Tapered Image RRL Properties of Multiple Detections\label{tab:rrl_notap_mult}}
\tablehead{
\colhead{Target} & \colhead{Field} & \colhead{Epoch\tablenotemark{a}}   & \colhead{\(\nu_L\)\tablenotemark{b}} & \colhead{$S_L$}          & \colhead{$V_{\rm LSR}$} & \colhead{$\Delta V$}    & \colhead{rms$_L$}      & \colhead{S/N} & \colhead{QF} \\
\colhead{}       & \colhead{}      & \colhead{}                         & \colhead{(MHz)}                      & \colhead{(mJy}           & \colhead{(km s$^{-1}$)} & \colhead{(km s$^{-1}$)} & \colhead{(mJy}         & \colhead{}    & \colhead{} \\
\colhead{}       & \colhead{}      & \colhead{}                         & \colhead{}                           & \colhead{beam\(^{-1}\))} & \colhead{}              & \colhead{}              & \colhead{beam$^{-1}$)} & \colhead{}    & \colhead{} 
}
\startdata
G289.944$-$00.889 & shrds191 & 2015 & 6800 & 8.60 $\pm$ 0.50 & 31.50 $\pm$ 0.70 & 23.60 $\pm$ 1.70 & 1.40 & 13.6 & B \\ 
G289.944$-$00.889 & g290.012-00.867 & 2014 & 5193 & 11.70 $\pm$ 3.10 & 31.80 $\pm$ 1.70 & 13.10 $\pm$ 4.20 & 5.10 & 3.7 & C \\ 
G290.012$-$00.867 & g290.012-00.867 & 2014 & 6134 &  \nodata & \nodata & \nodata & 3.00 & \nodata & D \\ 
G290.012$-$00.867 & g290.012-00.867 & 2014* & \nodata & 3.60 $\pm$ 2.50 & 14.70 $\pm$ 8.30 & 24.90 $\pm$ 8.60 & 1.80 & 4.5 & C \\ 
G290.323$-$02.984 & g290.323-02.984 & 2014 & 5874 &  \nodata & \nodata & \nodata & 2.50 & \nodata & D \\ 
G290.323$-$02.984 & g290.323-02.984 & 2014* & \nodata & 3.50 $\pm$ 2.10 & -17.70 $\pm$ 8.00 & 28.30 $\pm$ 8.30 & 1.50 & 5.6 & C \\ 
G290.385$-$01.042 & g290.385-01.042 & 2014 & 5911 &  \nodata & \nodata & \nodata & 2.30 & \nodata & D \\ 
G290.385$-$01.042 & g290.385-01.042 & 2014* & \nodata & 4.00 $\pm$ 2.90 & 9.91 $\pm$ 4.60 & 13.20 $\pm$ 4.70 & 1.90 & 3.4 & C \\
\enddata
\tablenotetext{a}{Rows with epochs including an asterisk (*) are copied directly from \citet{brown2017}. They did not give the weigted average frequency of their Hn\(\alpha\) RRL spectra.}
\tablenotetext{b}{This is the weighted average frequency of the \(<\)H\(n\alpha\)\(>\) stacked spectrum.}
\end{deluxetable*}
\end{longrotatetable}

\newpage

\begin{longrotatetable}
\begin{deluxetable*}{lcccrrrrrc}
\centering
\tablewidth{0pt}
\tabletypesize{\scriptsize}
\tablecaption{\textit{uv}-Tapered Image RRL Properties of Multiple Detections\label{tab:rrl_uvtap_mult}}
\tablehead{
\colhead{Target} & \colhead{Field} & \colhead{Epoch\tablenotemark{a}}   & \colhead{\(\nu_L\)\tablenotemark{b}} & \colhead{$S_L$}          & \colhead{$V_{\rm LSR}$} & \colhead{$\Delta V$}    & \colhead{rms$_L$}      & \colhead{S/N} & \colhead{QF} \\
\colhead{}       & \colhead{}      & \colhead{}                         & \colhead{(MHz)}                      & \colhead{(mJy}           & \colhead{(km s$^{-1}$)} & \colhead{(km s$^{-1}$)} & \colhead{(mJy}         & \colhead{}    & \colhead{} \\
\colhead{}       & \colhead{}      & \colhead{}                         & \colhead{}                           & \colhead{beam\(^{-1}\))} & \colhead{}              & \colhead{}              & \colhead{beam$^{-1}$)} & \colhead{}    & \colhead{} 
}
\startdata
G289.944$-$00.889 & shrds191 & 2015 & 6429 & 8.90 $\pm$ 0.80 & 32.30 $\pm$ 0.90 & 21.40 $\pm$ 2.10 & 1.70 & 11.0 & B \\ 
G289.944$-$00.889 & g290.012-00.867 & 2014 & 5276 &  \nodata & \nodata & \nodata & 4.40 & \nodata & D \\ 
G290.012$-$00.867 & g290.012-00.867 & 2014 & 6035 &  \nodata & \nodata & \nodata & 3.00 & \nodata & D \\ 
G290.012$-$00.867 & g290.012-00.867 & 2014* & \nodata & 3.60 $\pm$ 2.50 & 14.70 $\pm$ 8.30 & 24.90 $\pm$ 8.60 & 1.80 & 4.5 & C \\ 
G290.323$-$02.984 & g290.323-02.984 & 2014 & 5819 &  \nodata & \nodata & \nodata & 2.50 & \nodata & D \\ 
G290.323$-$02.984 & g290.323-02.984 & 2014* & \nodata & 3.50 $\pm$ 2.10 & -17.70 $\pm$ 8.00 & 28.30 $\pm$ 8.30 & 1.50 & 5.6 & C \\ 
G290.385$-$01.042 & g290.385-01.042 & 2014 & 5921 &  \nodata & \nodata & \nodata & 2.40 & \nodata & D \\ 
G290.385$-$01.042 & g290.385-01.042 & 2014* & \nodata & 4.00 $\pm$ 2.90 & 9.91 $\pm$ 4.60 & 13.20 $\pm$ 4.70 & 1.90 & 3.4 & C \\
\enddata
\tablenotetext{a}{Rows with epochs including an asterisk (*) are copied directly from \citet{brown2017}. They did not give the weigted average frequency of their Hn\(\alpha\) RRL spectra.}
\tablenotetext{b}{This is the weighted average frequency of the \(<\)H\(n\alpha\)\(>\) stacked spectrum.}
\end{deluxetable*}
\end{longrotatetable}

\end{document}